\documentclass[12pt,a4paper]{article}
\usepackage{amssymb}
\usepackage{rotating}
\usepackage{amsfonts}
\usepackage{amsmath}
\usepackage{amsthm}
\usepackage{caption}
\usepackage{float}
\usepackage{graphicx}
\usepackage[colorlinks=true, urlcolor=blue, pdfborder={0 0 0}]{hyperref}
\usepackage{wrapfig}
\usepackage{enumerate}
\usepackage{multicol}
\usepackage{srcltx}
\usepackage{array}
\newtheorem{theorem}{Theorem}[section]
\newtheorem{proposition}{Proposition}[section]
\newtheorem{lemma}{Lemma}[section]

\newtheorem{problem}{Problem}

\newtheorem{definition}{Definition}[section]

\newtheorem{remark}{Remark}
\newcommand{\wh}{\widehat}
\newcommand{\ov}{\overline}
\newcommand{\wt}{\widetilde}

\def\bbr{{\mathbb R}}

\def\text#1{\hbox{#1}}
\def\proof{{\noindent \bf Proof. }}

\def\endproof{\mbox{\ $\qed$}}

\def\E{{\bf E}}

\def\K{{\bf K}}
\def\P{{\bf P}}

\def\J{{\bf J}}

\def\ES{{\bf ES}}

\def\d{\mathrm{d}}
\def\build #1_#2{\mathrel{\mathop{\kern 0pt #1}\limits_\zs{#2}}}
\newcommand{\zs}[1]{{\mathchoice{#1}{#1}{\lower.25ex\hbox{$\scriptstyle#1$}}
{\lower0.25ex\hbox{$\scriptscriptstyle#1$}}}}

\numberwithin{equation}{section}

\def\proof{{\noindent \bf Proof. }}
\def\endproof{\mbox{\ $\qed$}}
\usepackage[top=1.5in, bottom=1.5in, left=1.1in, right=1in]{geometry}

\begin{document}

\title{Optimal investment and consumption with downside risk constraint in jump-diffusion models}

\author{ Thai Huu Nguyen\thanks {Correspondance email: thai.nguyen@univ-ulm.de; thaibopy@gmail.com
}
\\[3mm] {\small Department of Mathematics and Economics, Universit\"{a}t Ulm, Germany}
}

\vspace{2mm}
\date{\today}
\maketitle

\begin{abstract}
This paper extends the results of the article [C. Kl\"{u}ppelberg and S. M. Pergamenchtchikov. Optimal consumption and investment with bounded downside risk for power utility functions. In Optimality and Risk: {\it Modern Trends in Mathematical Finance. The Kabanov Festschrift}, pages 133-169, 2009] to a jump-diffusion setting. We show that under the assumption that only positive jumps in the asset prices are allowed, the explicit optimal strategy can be found in the subset of admissible strategies satisfying the same risk constraint as in the pure diffusion setting. When negative jumps probably happen, the regulator should be more conservative. In that case, we suggest to impose on the investor's portfolio a stricter constraint which depends on the probability of having negative jumps in the assets during the whole considered horizon.
\end{abstract}

\vspace{2mm}

\vspace{2mm}

\section{Introduction}\label{Intro}
One of the principal problems in mathematical finance is to consider a combination of optimal investment during a fixed investment interval $[0,T]$ and optimal terminal wealth at maturity. In particular, starting with an initial wealth $x$, the investor tries to maximize the following cost function 
\begin{equation}
J^\alpha(x):=\E_x \left(\int_0^T U_1(c_t)\d t+U_2(X_T^\alpha)\right),
\label{itro}
\end{equation}
where $U_1,U_2$ are two given utility functions, $c_t$ is the rate of consumption and $X_T^\alpha$ is the terminal wealth depending on the control process $\alpha$.  Such problems are of prime interest for institutional investors, selling asset funds
to their customers, who are entitled to certain payment during the duration of an
investment contract and expect a high return at maturity. In reality, financial activities must respect to some mandatory regulations mathematically defined by a risk measure \cite{Artzner99}. Note that Value-at-Risk (VaR) and Expected Shortfall (ES) are such measures endorsed by the Basel Committee on Banking Supervision. However, VaR is not a convex risk measure and only the probability to exceed a VaR bound is considered, not the values of the losses. It has ben shown that ES, defined as the conditional expectation of losses
above VaR, can be employed to fix this limitation. The literature for the problem of optimal portfolio under risk constraints is vast and we refer to \cite{Atkinson05,CuocoHuaIssaenko01, BasakShapiro01, Yiu04,EmmerKlupperbergKorn01}
and the references therein for more detailed discussions. 

Note that in order to satisfy the Basel committee requirements, investors must control the
level of loss throughout the investment horizon. This problem is studied by Kl\"{u}ppelberg and Pergamenshchikov \cite{KluppelbergPergamenchtchikov2009,kluppelberg2009b} for power and logarithm utility functions in the class of the nonrandom financial strategies with continuous asset dynamics. Chouaf and Pergamenschikov \cite{Chouaf-Perg} study the optimal investment problem with the
uniformly bounded VaR for random admissible financial
strategies.

Recent research in finance has paid attention to empirical evidence of jumps in stock returns \cite{Eraker03,Eraker04}. In fact, by incorporating jumps into the model we can allow to have sudden but infrequent market movements of large magnitude, and thus capture the skewed and fat-tailed features of stock return distributions. It has been shown by many empirical and theoretical studies
that the jump risk has a substantial impact on portfolio choice,
risk management and option pricing \cite{Merton,DuffiePan01}. In particular,  optimal portfolios held by an investor facing jump risks may significantly differ from those in the absence of jumps, and ignoring jumps may lead substantial losses \cite{DasUppal04}. Unlike pure-diffusion models with portfolio constraints, the martingale duality approach \cite{KaratzasShreve1998} may not be applied directly to a jump-diffusion model since the incompleteness caused by jumps in a jump-diffusion model may not be removed through the well-known completion techniques.

Let us only mention some among a vast recent literature of the problem of optimal portfolio choice with jumps. \cite{DuffiePan01} studies approximation for computing value at risk and other risk measures for portfolios that may include options and other derivatives with defaultable counterparties or borrowers. \cite {Framstad98,Framstad01} study the problem of optimal portfolio in a one-dimension jump-diffusion context with or without transaction costs. \cite{liu2003dynamic,DasUppal04, Aitsahalia09} solve the portfolio selection problems in jump-diffusion models with constant coefficients and no portfolio constraints where there is only one type of jumps. \cite{Jin13} considers the dynamic portfolio choice problem in a jump-diffusion model with constraints on portfolio weights using a particularly embedding the constrained problem in an appropriate family of unconstrained ones. 

	%
In this paper, we extend the results in \cite{KluppelbergPergamenchtchikov2009} to a jump-diffusion setting. Let us emphasize that even in the absence of risk constraint, problem \eqref{itro} in jump-diffusion models has not been well studied in the literature. In general, it is challenging to obtain the optimal solution in explicit form when investment and consumption are both considered. For the constrained problem, it is impossible to obtain an explicitly equivalent constraint on portfolio from the given VaR/ES constraint, which is imposed on the wealth process, due to the presence of jumps. As a result, the HJB approach may not be easily applied to solve the problem as e.g. in \cite{Yiu04}. 
The finding in this present paper is two-fold. First, we show that, under suitable choices for risk preference and for two identical power utility functions $U_1(x)=U_2(x)=x^{\gamma}$, the optimal solution of the unconstrained problem is still optimal for the constrained problem when jumps in the assets are non negative.  When $U_i(x)=x^{\gamma_i}$ with $0<\gamma_1\neq\gamma_2<1$, the impact of constraint is dramatic and it is optimal for the investor to consume all. Second, when negative jumps are allowed in the asset prices, we propose a slightly stricter constraint that takes into account the probability of having negative jumps in the horizon. Thus, the paper may give a reasonable choice to the regulator in designing regulatory policies for models with jumps. 

Let us shortly explain our main idea. First, from the regulator's point of view, jumps in the asset are not always unexpected. Roughly speaking, when jumps in asset are non negative (e.g. the markets are blooming), the risk of the investor's portfolio is less or equal to the risk in the absence of jumps if both are constructed optimally. The assumption of non negative jumps is positively correlated to the possibility of investing more in the risky asset. It is then reasonable to look for optimal strategies among those satisfying a constraint with the same confidence level imposed on a ''modified'' wealth process which is simply obtained by ignoring jumps in the initial wealth process. Because jumps are non negative, the corresponding admissible strategies form a subset of the initial admissible set. As explicitly shown in \cite{KluppelbergPergamenchtchikov2009}, this admissible subset can be deduced from an equivalent constraint which is directly imposed on the strategy. This then allows to employ the HJB approach or the direct method in \cite{KluppelbergPergamenchtchikov2009} to get the optimal solution. In such a jump-diffusion setting but with the same modified constraint as used in the pure diffusion case, the regulator now needs to check whether the constraint is fulfilled by the optimal solution of the unconstrained problem in the jump-diffusion model. This can be confirmed under some condition on the risk preference parameter. Intuitively, the regulator should be more conservative if  negative jumps probably happen. In that case, a slightly stricter constraint depending on the probability that there are negative in the risky assets during the investment horizon can be applied to ensure that the analysis for the case of non negative jump is still valid.

\vspace{2mm}
The remainder of the paper is organized as follows. Section \ref{Model} formulates the market model. We provide in Section \ref{Noconstr} a complete analysis for the unconstrained problem in jump-diffusion settings. The main results of the paper are reported in Section \ref{Constr}. Section \ref{General} considers the general case when the terminal utility and the consumption functions are different. Section \ref{Negative} studies the case where negative jumps are allowed. Auxiliary results are reported in Appendix.
\section{The market model}\label{Model}
Consider a financial market with $d$ risky assets  defined in the horizon $[0,T]$ by the system
\begin{align}
\d S^j_t=S^j_\zs{t^{-}} \left(\mu_t^j \d t+\sum_\zs{i}^d\sigma^{ij}_t \d W^j_t+\int_\zs{\bbr} z \wt{J}^j(\d t\times\d z)\right),\quad S^j_0=s^j>0,
\label{eq:Mol.1}
\end{align}
where
$W_t=(W^1_t,\dots,W^d_t)$ is a standard Brownian motion and $\wt{J}^j$ is the compensated random Poisson measure generated by the compound Poisson process 
$$
\zeta^j_t=\sum_{k=1}^{N^j_t} \xi_k^j,\quad j=1,\dots,d.
$$
The riskless asset is given by $\d S^0_t= r_t S^0_t,\,S^0_0=1,$ where $r_t$ is the riskless interest rate. We assume furthermore that $\zeta_j=(\zeta^1_t,\dots,\zeta^d_j)$ is of independent component vector process and independent of the Brownian motion $W_t$. Let 
$$\nu(\d z)\times\d t= (\nu^1(\d z^1),\dots, \nu^d(\d z^d))\times \d t
$$ be the $d$-dimension L\'evy measure of $\zeta_j$ then it is well-known that for any $j\in[1,\dots,d]$, $\nu^j(\d z^j)=\lambda^j F^j(\d z^j)$, where $F^j$ is the common distribution function of the jump sizes $(\xi_k^j)_\zs{k\ge 1}$ and $\lambda^j$ is the intensity of the Poisson process $N^j_t$. To guarantee the positivity of the stock prices, we assume that 
$$
\xi_k^j> -1, \quad \mbox{a.s. for any}\quad 1\le k
$$
and $\E[\xi_1^j]^2<\infty$ for all $1\le j \le d$.
Denote by $\mu_t=(\mu^1_t,\dots,\mu_t^d)$ the vector of stock appreciation and by $\sigma_t=(\sigma^{ij}_t)_\zs{1\le i,j\le d}$ the matrix of the stock volatilities. We suppose throughout the paper that these processes are deterministic, continuous and $\sigma_t$ is non-singular for Lebesgue almost surely $t\ge 0$.

Let $\phi_t=(\phi_t^0,\phi^1_t,\dots,\phi^d_t)$ be the amount of investment into bond and stocks at time $t\ge 0$. The wealth process is then given by
$$
X_t=\phi^0_t S^0_t+\sum_{j=1}^d \phi_t^j S^j_t.
$$
Assume moreover that consumption is possible and defined by the a progressively measurable non-negative process $c_t$ verifying 
$\int_0^T c_t\d t<\infty,\quad a.s.$. The strategy $\phi_t$ is called self-financing if the wealth process $X_t$ satisfies the following equation
\begin{equation}
\d X_t =\sum_{j=0}^d \phi^j_t \d S^j_t-c_t \d t,\quad X_0=x>0.
\label{eq:Mod.2}
\end{equation}
For $1\le j\le d$, denote by 
\begin{equation}
\pi^j_t=\frac{\phi^j_t S^j_t}{\sum_{j=0}^d \phi^j_t \d S^j_t}
\label{eq:Mod.3}
\end{equation}
the fraction of the wealth invested into $j$-th asset. The portfolio process $\pi_t=(\pi_t^1,\dots,\pi_t^d)$ is assumed to be \emph{cadlag} and $\int_0^T \left\|\pi\right \|_t<\infty$ almost surely. For convenience, define 
\begin{equation}
y_t=(y_t^1,\dots,y_t^d):=\sigma_t' \pi_t \quad \mbox{and}\quad \theta_t=(\theta_t^1,\dots,\theta_t^d):=\sigma_t^{-1}(\mu_t-r_t{\bf 1}).
\label{eq:Mol.4}
\end{equation}
With these notations, we rewrite \eqref{eq:Mod.2} as 
\begin{equation}
\d X_t =X_t(r_t+y'_t\theta_t)\d t-c_t \d t+X_t y_t^{'}\d W_t+X_\zs{t^{-}} \int_{\bbr^d}\pi_t^{'} z \d\wt{J}(\d z\times \d t), \quad X_0=x>0,
\label{eq:Mod.5}
\end{equation}
where $\wt{J}(\d z\times \d t):=(\wt{J}^1(\d z^1\times \d t),\dots, \wt{J}^d(\d z^d\times \d t))$.
Suppose that consumption is a proportion of wealth, i.e. $c_t=v_t X_t$, where $v_t$ is a non-negative deterministic process holding $\int_0^T v_t \d t<\infty$. Define $\alpha_t=(y_t,v_tX_t)$ and use the notation $X^\alpha$ to emphasize that the wealth process is defined with some control $\alpha$. Now, \eqref{eq:Mod.5} can be rewritten as 
\begin{equation}
\d X_t^\alpha =X_t^\alpha(r_t+y'_t\theta_t-v_t)\d t+X_t^\alpha y_t^{'}\d W_t+X_\zs{t^{-}}^\alpha \int_{\bbr^d}\pi_t^{'} z \wt{J}(\d z\times \d t), \quad X_0^\alpha=x>0.
\label{eq:Mod.6}
\end{equation}
Denote by  
\begin{equation}
{\cal E}_t(y)=\exp\left\{\int_0^t y'_s\d W_s -\frac{1}{2}\int_0^t\vert y\vert^2_s \d s\right\}
\label{eq:Mol.6-1}
\end{equation}
the stochastic exponential and put
\begin{equation}
\wh{\theta}_t:=\sigma_t^{-1}(\mu_t-r_t{\bf 1}- {\xi}_{\lambda})\, \quad \mbox{with}\quad  {\xi}_{\lambda}=(\lambda^1\E\xi^1_1, \lambda^2\E\xi^2_1,\dots,\lambda^d\E\xi^d_1).
\label{eq:Mol.6-2}
\end{equation}
Then, by It\^o's formula for jump processes, it is straightforward to see that \eqref{eq:Mod.6} admits the following solution
\begin{equation}
X_t^\alpha =xe^{R_t-V_t+(y,\wh{\theta})_t} {\cal E}_t(y) P_t^\pi(\xi),
\label{eq:Mol.7}
\end{equation}
where 
\begin{equation}
R_t=\int_0^t r_s\d s,\quad V_t=\int_0^t v_s \d s, \quad (y,\wh{\theta})_t=\int_0^t y'_s \wh{\theta}_s \d s
\label{eq:Mol.8}
\end{equation}
and the jump part $P_t^{\pi}(\xi)$ defined as
\begin{equation}
P_t^{\pi}(\xi)=\exp\left\{\sum_{j=1}^d\int_0^t\int_{\bbr}\ln(1+\pi^j_sz^j) {J}^j(\d z^j\times \d t)\right\}=\prod_{j=1}^d \prod_{k=1}^{N_t^j} (1+\pi^j_\zs{\tau_k^{j{-}}} \xi^j_k),
\label{eq:Mol.9}
\end{equation}
where $(\tau^j_k)_\zs{k\ge1}$ is the sequence of jump times of $N_t^j$, $1\le j\le d.$
Admissible strategies are specified as follows.
\begin{definition}
The process $\alpha=(\alpha_t)_\zs{0\le t\le T}$ is called \emph{admissible} if 
\begin{enumerate}
	\item $y_t$ and $c_t$ are predictable,
	\item $c_t(\omega)\ge 0$, for a.e. $(t,\omega),$
	\item for any $1\le j\le d,$ $\pi_t^j(\omega)\in[0,1] $ for a.e. $(t,\omega),$ 
	\item equation \eqref{eq:Mod.6} admits a unique strong solution $X_t^\alpha$ defined as in \eqref{eq:Mol.7}.
\end{enumerate}
\end{definition}
Note that the third condition, interpreted as the short selling prohibition, is necessary to make sure that $X_t^\alpha$ is positive. We denote by ${\cal D}$ the class of all admissible control processes. The effect of no short selling has been investigated for two CRRA utility functions in constant coefficient markets is studied in \cite{Jin2013}.
\begin{remark}
In \cite{Jin2013}, the authors study dynamic optimal portfolio choice in a jump-diffusion with investment constraint including no-short selling and no-borrow constraints.  The key idea is to construct a set of fictitious markets by adjusting the interest rate and the drift terms of stock prices. The constrained consumption-investment problem in the original market is converted into an unconstrained one in a set of fictitious markets. Then, the optimality for the original market with investment constraints can be obtained by optimally adjusting the interest rate and the stock price drift terms in the fictitious markets. For detailed discussions, see \cite{Jin2013}.
\end{remark}

Assume now that the investor wants to optimize his expected utility of consumption over the time interval $[0,T]$ and his wealth at terminal horizon. In other words, for an initial endowment $x>0$ and a control process $\alpha_t\in {\cal U}$, we consider the cost function of the form
\begin{equation*}
J^\alpha(x):=\E_x \left(\int_0^T U_1(c_t)\d t+U_2(X_T^\alpha)\right),
\label{eq:Mol.10.0}
\end{equation*}
where $U_i, i=1,2$ are utility functions and $\E_x$ is the expectation operator conditional on $X_0^\alpha=x$. For power utility functions problems, we choose $U_i(u)=u^{\gamma_i},$ for $u\ge 0$, with $0<\gamma_i\le 1, i=1,2$ and the cost function is then given by
\begin{equation}
J^\alpha(x):=\E_x \left(\int_0^T c^{\gamma_1}_t\d t+(X_T^\alpha)^{\gamma_2}\right).
\label{eq:Mol.10}
\end{equation}

Let us now precise the risk constraints considered in this paper. 
The VaR defined below is also known as ''Captial at Risk''. Here we adopt the idea in \cite{EmmerKlupperbergKorn01, KluppelbergPergamenchtchikov2009}.

\begin{definition}[Value at Risk]\label{Def.1}
For an initial endowment $x>0$, a control $\alpha$ and a real number $0<\beta\le 1/2$, we define the \emph{Value at Risk} (VaR) of the wealth process $X_t^\alpha$ as
\begin{equation}
VaR_t(x,\alpha,\beta):=xe^{R_t}-q_\beta(X^\alpha_t),
\label{eq:Mol.11}
\end{equation}
where $q_\beta(X^\alpha_t)$ is the lower $\beta$-quantile of $X^\alpha_t$.
\end{definition}
For definition of lower quantile we refer to Definition \ref{Def.0-1}. It should be stressed that the above definition is consistent with the setting in the well-known paper \cite{BasakShapiro01} with limit loss $L_t=(1-\kappa) xe^{R_t}$, which needs to be checked dynamically at any time $t\in[0,T]$.
Note that $q_\beta<0$ for $0<\beta<1/2$. The level of risk is characterized by $\kappa x e^{R_t}$ for some coefficient $0<\kappa<1$, which represents a liability level for the investor's portfolio. Now, for some $\kappa$, we look for a strategy  $\alpha\in{\cal D}$ for which the Value at Risk is uniformly bounded by $\kappa x e^{R_t}$, i.e. we are working under the following dynamical risk constraint
\begin{equation}
\sup_\zs{0\le t\le T} \frac{VaR_t(x,\alpha,\beta)}{\kappa xe^{R_t}}\le 1.
\label{eq:Mol.13}
\end{equation}
Also remark that the risk level $\beta$ and $\kappa$ are determined by the regulator. In some cases, the investor can take $\kappa$ in a given range. If $\kappa$ is close to 0, risk of the portfolio risk is at low level whereas $\kappa$ is near 1, the portfolio has a high risk of loss as in the unconstrained problem. In the latter case, the risk limit may not be active. 
Let us now define the second kind of constraint which, unlike the quantile, focuses on the averaged value of loss.

\begin{definition}
For an initial endowment $x>0$, a control $\alpha$ and a real number $0<\beta\le 1/2$, we define the \emph{Expected Shortfall} (\ES) of the wealth process $X_t^\alpha$ as
\begin{equation}
\ES_t(x,\alpha,\beta):=xe^{R_t}-\ES_\beta(X^\alpha_t),
\label{eq:Mol.14}
\end{equation}
where $\ES_\beta(X^\alpha_t)$ is the classical expected shortfall.
\end{definition}
For the reader's convenience, the classical definition of expected shortfall is given in Definition \ref{Def.2}. Note that the investor's portfolio can be controlled by imposing continuously the following constraint:  ${\ES_t(x,\alpha,\beta)}\le {\kappa xe^{R_t}} $ for all $t\in[0,T]$. Then, the same interpretation can be observed as in the case of VaR constraint. In Section \ref{Constr}, we study the following constrained problems:
\begin{problem}\label{Prob.2} Given $x>0$ and $0<\kappa<1$, find strategy $\alpha^*\in {\cal D}$ which solves 
$$
\max_\zs{\alpha\in {\cal D}} J^{\alpha}(x)\quad \mbox{\normalfont{subject to} }\quad \sup_\zs{0\le t\le T} \frac{VaR_t(x,\alpha,\beta)}{\kappa xe^{R_t}}\le 1;
$$
\end{problem}
\begin{problem}\label{Prob.3}Given $x>0$ and $0<\kappa<1$, find strategy $\alpha^*\in {\cal D}$ which solves 
$$
\max_\zs{\alpha\in {\cal D}} J^{\alpha}(x)\quad \mbox{\normalfont{subject to} }\quad \sup_\zs{0\le t\le T} \frac{\ES_t(x,\alpha,\beta)}{\kappa xe^{R_t}}\le 1.
$$
\end{problem} 
Let us make here some comments on the above problems. First, the classical martingale method \cite{KaratzasShreve1998} seems to be impossible for such problems with dynamic risk constraint. Note that in pure diffusion models, it is possible to transform the VaR/ES constraint into a so-called portfolio constraint, i.e. constraint on strategies, then martingale duality might be applied by considering a new artificial market as in Cvitanic and Karatzas \cite{cvitanic1992convex,cvitanic1993}. However, this transformation seems highly challenging in the presence of jumps. Another possibility is to solve the problem approximately, i.e. we first approximate the risk constraint by a portfolio constraint then
consider the corresponding approximate HJB. It is likely that such procedure need to be done in a delicate asymptotic analysis. Lastly, it might be possible to employ the so-called weak dynamic programming principle for state constraint suggested by Bouchard and Nutz  \cite{bouchard2012weak}, but then we need to look for optimal solution in the sense of viscosity solution.

Below, wishing to get a closed form of the optimal solution, we adapt the direct
method used by Kluppelberg and Pergamenschikov \cite{KluppelbergPergamenchtchikov2009}.

\section{Optimization problem without constraints}\label{Noconstr}
We provide a detailed analysis for the unconstrained problem
\begin{problem}\label{Prob.1} Given $x>0$ and $0<\kappa<1$, find strategy $\alpha^*\in {\cal D}$ which solves 
$$
\max_\zs{\alpha\in {\cal D}} J^{\alpha}(x),
$$
where $ J^{\alpha}(x)$ is the cost function defined in \eqref{eq:Mol.10}.
\end{problem}
Note that there have been very few studies in jump settings where both optimal investment and consumption are combined.  Although our main aim is to deal with the constrained problem, this section may be seen as another contribution of the paper. First, the indirect value function is given by
\begin{align}
u(t,x)&:=\E \left(\int_t^T c^{\gamma_1}_s\d s+(X_T^\alpha)^{\gamma_2}\big\vert X_t^\alpha=x \right).
\end{align}
For completeness, we begin with the simplest case $\gamma_1=\gamma_2=1$ thought this case has 
economically not much sense without a risk constraint. A detailed proof is given in Appendix \ref{App:1}.
\begin{theorem}\label{Th.Noconstr.1}
Consider Problem \ref{Prob.1} with $\gamma_1=\gamma_2=1$. Assume that $\mu^j_t\ge r_t,$ for all $1\le j\le d$ and $t\in[0,T]$. Then,
\begin{enumerate}
 \item If $\Vert \mu-r{\bf 1}\Vert_\zs{T}=0$ then any control $\alpha^*_t=(\pi^*_t,0)$ with $\pi^{*j}\in[0,1]$ is an optimal solution and the corresponding optimal value of $J^\alpha(x)$ is given by $J^*(x)=xe^{R_T}$.
\item If $\Vert \mu-r{\bf 1}\Vert_\zs{T}>0$ then the control $\alpha^*_t=(\pi^*_t,0)$ with 
\begin{equation}
\pi^*_t=(\mu_t-r_t{\bf 1})\sqrt{T}\Vert \mu-r{\bf 1}\Vert_\zs{T}^{-1}
\label{eq:No.constr.0-0}
\end{equation}
is the optimal solution and the corresponding optimal value of $J^\alpha(x)$ is given by $J^{*}(x)=xe^{R_T+\sqrt{T}\Vert \mu-r{\bf 1}\Vert_\zs{T}}$.
\end{enumerate}
\end{theorem}


\subsection{The case $\gamma_1=\gamma_2=\gamma \in(0,1)$}\label{sec:equalgamma}
We now study the unconstrained problem for the case $\gamma_1=\gamma_2=\gamma \in(0,1)$. Let us first compute the value function. By \eqref{eq:Mol.7} and noting that $\E{\cal E}_t(\gamma y)=1$ one gets
\begin{equation}
\E (X_t^\alpha)^\gamma=x^\gamma e^{\gamma (R_t-V_t+(y,\wh{\theta})_t-\frac{1-\gamma}{2}\Vert y\Vert_t^2)}\E P_t^\pi(\gamma\xi).
\label{eq:Noconstr.02}
\end{equation}
On the other hand, by Lemma \ref{Le:Levy}
\begin{align}
\E P_t^\pi(\gamma\xi)&=\E \exp\left\{\int_0^t\int_{\bbr^{d}} \ln(1+\pi_s z)^\gamma J(\d z\times\d s)\right\}\notag\\
&=\E \exp\left\{\int_0^t\int_{\bbr^{d}} ((1+\pi_s z)^\gamma-1) \nu(\d z\times\d s)\right\}.
\label{eq:}
\end{align}
Therefore,
\begin{equation}
\E (X_t^\alpha)^\gamma=x^\gamma\exp\left\{\gamma D_t+\int_0^t\int_{\bbr^{d}} Q_s^\pi(z) \nu(\d z\times\d s)\right\},
\label{eq:Noconstr.03}
\end{equation}
and
\begin{equation}
D_t=R_t-V_t+(y,{\theta})_t-\frac{1-\gamma}{2}\Vert y\Vert_t^2, \quad Q_t^\pi(z)=(1+\pi_t z)^\gamma-\gamma\pi_t z-{\bf 1}.
\label{eq:}
\end{equation}
Hence, the value function is given by 
\begin{equation}
J^\alpha(x)=x^\gamma \left(\int_0^T e^{\gamma D_t+ \int_0^t\int_{\bbr^{d}} Q_t^\pi(z) \nu(\d z\times\d t)} v^\gamma_t \d t +e^{\gamma D_T+ \int_0^T\int_{\bbr^{d}} Q_t^\pi(z) \nu(\d z\times\d t)}\right).
\label{eq:}
\end{equation}
From the dynamics of $X_t^\alpha$ given in \eqref{eq:Mod.6}, one gets the HJB equation for the unconstrained problem
\begin{equation}
\partial_t u(t,x)+\sup_\zs{\alpha\in {\cal D}} \left\{ {\cal A} ^\alpha u(t,x)+ x^\gamma v_t^\gamma\right\}=0, \quad u(T,x)=x^\gamma,
\label{eq:HJB}
\end{equation}
where the generator ${\cal A}^\alpha$ is defined by
\begin{align}
{\cal A}^\alpha u(t,x)&=x(r_t+y'_t \wh{\theta}_t-v_t)\partial_xu(t,x)+\frac{1}{2}x^2 y_t y'_t\partial_{xx}^2u(t,x)\notag\\
&+\sum_\zs{j=1}^d \int_\bbr (u(t,x+x\pi^j_tz)-u(t,x)-x\pi^j_tz\partial_xu(t,x))\nu^j(\d z),
\label{eq:Oper}
\end{align}
where $y'_t \wh{\theta}_t=\sum_\zs{j=1}^d y^j_t \wh{\theta}^j_t$ is the scalar product.
The optimal necessary condition w.r.t $v$ is given by
$$
-xu_x+x^\gamma v^{\gamma-1}=0 \Longleftrightarrow  c=\left(\frac{u_x}{\gamma}\right)^{\frac{1}{\gamma-1}}.
$$
We try to find a solution of the form $u(t,x)=\rho x^\gamma$ (i.e. $c=x\rho^{\frac{1}{\gamma-1}}$), where $\rho$ is a $t$-function to be determined. We have
$$
u_x=\gamma\rho x^{\gamma-1},\quad u_\zs{xx}=\gamma (\gamma-1)\rho x^{\gamma-2}.
$$
Substituting these formulas into \eqref{eq:HJB} we obtain 
\begin{align}
\rho\sup_\zs{\alpha\in {\cal D}} \left\{\gamma (r_t+y'_t \wh{\theta}_t)+\frac{1}{2} y_t^2 \gamma (\gamma-1)
+\sum_\zs{j=1}^d \K^j(\pi^j_t) \right\}+\rho'+ (1-\gamma)\rho^{\frac{\gamma}{\gamma-1}}=0,
\label{eq:HJB4-0}
\end{align}
where
\begin{equation}
\K^j(\pi):=\int_\bbr [(1+\pi z)^\gamma-1-\gamma\pi z]\nu^j(\d z).
\label{eq:K}
\end{equation}
Let $\sigma_t^{-1}=( \epsilon_{ij}(t))_\zs{d\times d}$ and $\partial \pi^j/ \partial {y_i}= \epsilon_{ij}(t)$ and $q=1/(1-\gamma)$. Now, by the necessary optimal condition in $y$ one has
\begin{equation}
\wh{\theta}_t^i +(\gamma-1)y_t^i+\sum_\zs{j=1}^d  \epsilon_{ij}(t) Q^{j}(\pi^j_t)=0,\quad y^i\ge 0\;, i=\overline{1,d},
\label{eq:HJB3}
\end{equation}
where 
\begin{equation}
Q^{j}(\pi)=\int_\bbr [(1+\pi z)^{\gamma-1}-1]z\nu^j(\d z).
\label{eq:HJB4}
\end{equation}
The next step is to solve the system \eqref{eq:HJB3}. Observe first that the functions 
$\K^j(\pi), j=1,\dots,d$ are concave on $[0,1]$, hence supremum in \eqref{eq:HJB4-0} is attainable and unique. For the moment, let us assume that $y^*\ge 0$ is a solution of system \eqref{eq:HJB3} and let $h^*$ be the corresponding supremum in \eqref{eq:HJB4-0}. We then get the following Bernoulli equation
\begin{equation}
\rho'(t)+ h^*(t)\rho(t)=(\gamma-1)\rho^{\frac{\gamma}{\gamma-1}}(t)\quad \mbox{with terminal condition}\quad \rho(T)=1.
\label{eq:Bernoulli-1}
\end{equation}
The solution of \eqref{eq:Bernoulli-1} is given by
\begin{equation}
\rho(t)=\left[\frac{g^q(T)+\int_t^T g^q(s)\d s}{g^q(t)}\right]^{1/q}, \quad \text{where}\quad g(t)=e^{\int_0^t h^*(s) \d s }.
\label{eq:rho}
\end{equation}
It follows that the optimal rule is given by $y^*_t$ (which is the solution of \eqref{eq:HJB3}) and 
\begin{equation}
v^*_t=\frac{g^q(t)}{g^q(T)+\int_t^T g^q(s)\d s}
\label{eq:HJB5}
\end{equation}
and the optimal value of value function can be determined. Notice that when there is no jumps in the asset prices ($\nu^j=0, \,i=\overline{1,d}$), we obtain
$h^*=\exp\{\gamma R_t+\frac{q-1}{2}\Vert \theta\Vert_t^2\}$ and the optimal value of $J$ is given by
$$J^*(x)=\max_\zs{\alpha\in {\cal D}}  J(x,\alpha)=J(x,\alpha^*)=x^\gamma \left(\Vert g\Vert ^q_\zs{q,T}+g^q(T)\right)^{1/q},$$
where $\Vert.\Vert_\zs{q,T}$ is defined by
\begin{equation}
\Vert f\Vert_\zs{q,T}=\left(\int_0^T\vert f\vert^q\d t\right)^{1/q},
\label{eq:HJB.5-1}
\end{equation}
which means that the result in \cite{KluppelbergPergamenchtchikov2009} is recovered.

We summarize the above analysis in the following statement.
\begin{theorem}\label{Th.Noconstr.2}
Assume that $y^*$ is a solution of system \eqref{eq:HJB3} and let $h^*$ be the corresponding supremum in \eqref{eq:HJB4-0}. Then, the optimal rule for the problem $
\max_\zs{\alpha\in {\cal D}} J^{\alpha}(x)
$ is given by $\alpha^*=(y^*,v^*)$ with $v^*$ defined by \eqref{eq:HJB5}. The wealth process is given by
\begin{equation}
\d X_t^* =X_t^*(r_t+y_t^*\theta_t)\d t-c_t^* \d t+X_t^* y_t^*\d W_t+X_\zs{t^{-}}^* \int_{\bbr^d}\pi_t^* z \d\wt{J}(\d z\times \d t), \quad X_0^*=x>0.
\label{eq:HJB.7}
\end{equation}
\end{theorem}
\proof It it not difficult to check that all necessary conditions for the usual verification theorem (see e.g. \cite{Touzi2004, OksendalSulem07}) are satisfied. \endproof

\begin{remark}
It is instructive to verify the optimality by using the martingale optimality principle: the value function is a supermartingale for any admissible strategy but it becomes a martingale under the optimal strategy. This can be checked by applying It\^o's formula for the process $\rho(t) (X_t^*)^{\gamma}$, where
$\rho(t)$ is given by \eqref{eq:rho}.
\end{remark}

\begin{remark}
The case $\gamma_1\neq\gamma_2$ is more challenging to show since the function $\rho$ in \eqref{eq:rho} should be chosen as an appropriate combination of two functions $g^{q_i}$, $i=1,2$. This could be done using a similar argument of Theorem 2 in \cite{KluppelbergPergamenchtchikov2009}. However, it is also possible to see that the corresponding optimal solution does not satisfy the VaR/ES risk constraint. We will show in the next section that the presence of risk constraint has a strong impact on the investor' portfolio so that it is optimal for him to consume all. For that reason, we do not provide a detailed result for the unconstrained in this case and refer to  \cite{KluppelbergPergamenchtchikov2009} for a detailed analysis in the pure diffusion case.
\end{remark}

\subsection{One-dimension case}
We provide more analysis about the optimal rule for the case of one dimension. 
\begin{theorem}
Assume that for almost surely $t\in[0,1]$,
$$
\mu_t-r_t- {\xi}_{\lambda}>0, \quad \mu_t-r_t- {\xi}_{\lambda} +(\gamma-1)\sigma^2_t + \int_\bbr [(1+z)^{\gamma-1}-1]z\nu(\d z)<0.
$$
Then there exists a solution $\pi^*\in [0,1]$ to \eqref{eq:One1}. Let $G^*=G(\pi^*)$ be the optimal value of the function $G$ in the HJB equation \eqref{eq:One0} and consider $v^*$ defined by \eqref{eq:HJB5} in which we have replaced $h^*$ with $G^*$. Then, $(\pi^*,v^*)$ is an optimal solution to the problem $
\max_\zs{\alpha\in {\cal D}} J^{\alpha}(x)$ and the wealth process is given as the unique solution of
\begin{equation}
\d X_t^* =X_t^*(r_t+y_t^*\theta_t)\d t-c_t^* \d t+X_t^* y_t^*\d W_t+X_\zs{t^{-}}^* \int_{\bbr}\pi_t^* z \d\wt{J}(\d z\times \d t), \quad X_0^*=x>0.
\label{eq:HJB.7}
\end{equation}
\end{theorem}
\proof
As above, we get corresponding HJB equation
\begin{equation}
\rho'+ (1-\gamma)\rho^{\frac{\gamma}{\gamma-1}}+\rho \sup_\zs{\pi_t\in [0,1]} G(t,\pi_t)=0,
\label{eq:One0}
\end{equation}
where 
$$
G(t,\pi):=\gamma r_t+\gamma(\mu_t-r_t- {\xi}_{\lambda})\pi_t +\frac{1}{2}\sigma_t^2 \pi^2_t \gamma (\gamma-1)+\K(\pi_t),
$$
where $\K(\pi_t)$ is defined in \eqref{eq:K}.
The necessary condition for optimality is given by $ \eta(\pi_t)=0$, where
\begin{equation}
 \eta(\pi)=\partial_\pi G(t,\pi):=\mu_t-r_t- {\xi}_{\lambda} +(\gamma-1)\sigma^2_t \pi_t+Q(\pi_t),
\label{eq:One1}
\end{equation}
where $Q(\pi_t)$ defined by \eqref{eq:HJB4}. 
Note that $ \eta$ is continuous on $[0,1]$ with $ \eta(0)=\mu_t-r_t- {\xi}_{\lambda}>0$ and 
$$
 \eta(1)=\mu_t-r_t- {\xi}_{\lambda} +(\gamma-1)\sigma^2_t + \int_\bbr [(1+z)^{\gamma-1}-1]z\nu(\d z)<0.
$$
Furthermore, 
$$
 \eta'(\pi)=(\gamma-1)[\sigma^2_t+ \int_\bbr [(1+\pi z)^{\gamma-2}z^2\nu(\d z)]<0 
$$
since $\pi\in[0,1]$ and the support of $\nu$ is $(-1,\infty)$. So, by the theorem of mean values, for any $t\in [0,T]$, there exists $\pi_t^*$ such that $\partial_\pi G(t,\pi_t^*)=0$. The conclusion then follows the usual verification step. \endproof

\begin{remark}
For infinite horizon cases, extra conditions need to be imposed to get the uniform integrability. For general cadl\`ag coefficients, we can formulate a specific verification theorem as in \cite{KluppelbergPergamenchtchikov2009}.
\end{remark}
We now compare the optimal rule obtained above with the no-jump optimal strategy $(\bar{\pi}^*, \bar{v}^*)$ of the classical Merton problem.
\begin{lemma}[comparision]\label{Le:compare}
The presence of jumps reduces the quantity asset and consume more, i.e
$$
{v}^*_t\ge \bar{v}^*_t \quad \mbox{and}\quad {\pi}^*_t\le \bar{\pi}^*_t.
$$
\end{lemma}
 \proof A proof is given in Appendix \ref{App:com}.\endproof
\begin{figure}[H]%
\begin{center}
\includegraphics[width=0.7\columnwidth,height=7cm]{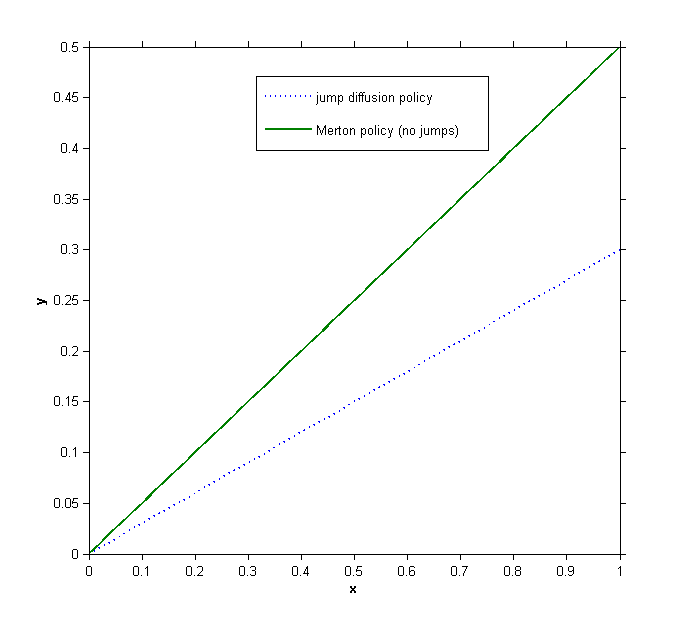}%
\caption{Optimal policy for the jump-diffusion and the pure diffusion markets}%
\label{}%
\end{center}
\end{figure}

\section{The optimization problems with risk constraint}\label{Constr}
In this section, we study the Problem \ref{Prob.2} and Problem \ref{Prob.3}. We assume in this section that the following condition holds:
\vspace{2mm}

\noindent{\bf Assumption} $(\J)$:{\em The jump sizes of stock prices are non-negative.}

\vspace{2mm}
\noindent As discussed in Section \ref{Intro}, when jumps in the assets are non negative (e.g. the markets are blooming), the risk of the investor's portfolio is smaller or at the same level than in the absence of jumps. Intuitively, positive jumps encourage the agent to invest more in the risky asset. It is then reasonable to look for the optimal strategy satisfying the constraint with the same confidence level but with ignored jumps. Because jumps are non negative, these strategies form a subset of the initial admissible set. As explicitly shown in \cite{KluppelbergPergamenchtchikov2009}, this admissible subset can be deduced from an equivalent constraint which is directly imposed on the strategy. This then allows to employ the HJB approach or the direct method in \cite{KluppelbergPergamenchtchikov2009} to get the optimal solution. 
%
		%

\subsection{The case $\gamma_1=\gamma_2=1$}
To present the results for the Problem \ref{Prob.2} with VaR constraint, we define $K_t=(\theta,\xi_{\lambda})_t \Vert \theta\Vert_\zs{T}^{-1}$ and
\begin{align}\label{eq.Constr.1}
{\rho}^*_\zs{VaR}:=\sqrt{(\Vert {\theta}\Vert_T-\vert q_\beta\vert-K_T)^2-2\ln(1-\kappa)}+\Vert {\theta}\Vert_T-K_T-\vert q_\beta\vert.
\end{align}

\begin{theorem}\label{Th.VaR.1}
 Consider Problem \ref{Prob.2} with $\gamma_1=\gamma_2=1$ under Assumption $(\J)$.
\begin{enumerate}
 \item If $\Vert \theta\Vert_\zs{T}=0$ then any control $\alpha^*_t=(\pi^*_t,0)$ with positive component vector $\pi^*_t$ satisfying 
$$
\Vert y^*\Vert_T\le \min\left(\sqrt{T}\Vert \sigma\Vert_T, \sqrt{(\vert q_\beta\vert+\Vert \wt {\xi}_{\sigma}\Vert_T)^2-2\ln(1-\kappa)}-\vert q_\beta\vert-\Vert \wt {\xi}_{\sigma}\Vert_T\right),
$$ with $ \wt {\xi}_{\sigma}=\xi_{\lambda}\sigma^{-1}$, is an optimal solution and the corresponding optimal value of $J^\alpha(x)$ is given by $J^{\alpha^*}(x)=xe^{R_T}$.

\item Suppose that $\wh{\theta}_t$ has non-negative components, i.e.
$$
\wh{\theta}^j\ge 0\quad \mbox{for all} \quad 1\le j\le d
$$
and $ \Vert{\xi}_{\lambda}\Vert$ defined in \eqref{eq:Mol.6-2} is strictly positive. Then, for 
\begin{equation}
\max(0, 1-e^{q_\beta^2/2-\vert q_\beta\vert \Vert \theta\Vert_T })<\kappa<1,
\label{eq:Constr.2}
\end{equation}
the control $(y^*_t,v^*_t)$ defined by
\begin{equation}
y^*_t={\theta}_t \Vert \theta\Vert_\zs{T}^{-1}\ov{\rho}^*_\zs{VaR},\quad v^*_t=0
\label{eq:Constr.3}
\end{equation}
is the optimal and $J^{\alpha^*}(x)=xe^{R_T+\Vert \theta\Vert_\zs{T}\ov{\rho}^*_\zs{VaR}}$ is the corresponding optimal value.
\end{enumerate}
\end{theorem}

\proof From \eqref{eq:Noconstr.1} one gets an upper bound for the value function 
$$
J^\alpha(x)\le xe^{R_T-V_T+(y,{\theta})_T}\le xe^{R_T+(y,{\theta})_T}\le xe^{R_T+\Vert y\Vert_T\Vert{\theta}\Vert_T}.
$$
We show that this upper bound can be attained by choosing the suitable optimal strategy. 

\noindent {\bf Step 1}: By the linear property of lower quantile one easily observes that the constraint in Problem \ref{Prob.2} is equivalent to 
\begin{equation}
\inf_\zs{0\le t\le T} q_{\beta}({\cal E}_t(y) P_t^\pi(\xi)) \ge e^{V_t-(y,\wh{\theta})_t}(1-\kappa).
\label{eq:Constr.4}
\end{equation}
Under assumption $(\J)$, the process $P_t^\pi(\xi)$ is bigger than 1 a.s. hence, by Lemma \ref{Le.Mol.2} $q_{\beta}({\cal E}_t(y) P_t^\pi(\xi))\ge q_{\beta}({\cal E}_t(y))$. Therefore, if $\alpha^*$ is an optimal solution to Problem \ref{Prob.2} under the weaker constraint 
\begin{equation}
\inf_\zs{0\le t\le T} q_{\beta}({\cal E}_t(y)) \ge e^{V_t-(y,\wh{\theta})_t}(1-\kappa),
\label{eq:Constr.5}
\end{equation}
then it is also an optimal solution to Problem \ref{Prob.2} with initial constraint \eqref{eq:Constr.4}. We now solve Problem \ref{Prob.2} with constraint \eqref{eq:Constr.5}, which is now can be transformed into an more explicit form 
\begin{equation}
\inf_\zs{0\le t\le T} \left\{-\frac{1}{2}\Vert y\Vert_t^2+q_{\beta}\Vert y\Vert_t-V_t+(y,\wh{\theta})_t\right\}\ge \ln(1-\kappa).
\label{eq:Constr.6}
\end{equation}

This is exactly the constraint in the diffusion case considered in \cite{KluppelbergPergamenchtchikov2009,kluppelberg2009b}.

\noindent{\bf Step 2}: Suppose first that $\Vert \theta\Vert_T=0$ hence $\theta_t=0$ for all $0\le t\le T$. Now, \eqref{eq:Constr.6} becomes
\begin{equation}
\inf_\zs{0\le t\le T} \left\{-\frac{1}{2}\Vert y\Vert_t^2+q_{\beta}\Vert y\Vert_t-V_t-(y,\xi_{\lambda})_t\right\}\ge \ln(1-\kappa),
\label{eq:Constr.7}
\end{equation}
which is satisfied if 
$$
 -\frac{1}{2}\Vert y\Vert_T^2+q_{\beta}\Vert y\Vert_T-V_T-\Vert y\Vert_T \Vert\xi_{\lambda}\Vert_T\ge \ln(1-\kappa)
$$
since $q_\beta<0$. The latter inequation has solution $\Vert y\Vert_T \in[0, \rho_0]$, where
\begin{equation}
\rho_0=\sqrt{(\vert q_{\beta}\vert+\Vert\xi_{\lambda}\Vert_T)^2-2\ln(1-\kappa)}-\vert q_{\beta}\vert-\Vert\xi_{\lambda}\Vert_T.
\label{eq:Constr.8}
\end{equation}
Thus, for $\alpha^*=(\pi^*_t,0)$, with any non-negative component vector $\pi^*_t=y_t^*\sigma^{-1}_t$ and $y^*_t$ satisfying $\Vert y\Vert_\zs{T}\le \min({\sqrt{T}\Vert \sigma\Vert_T, \rho_0})$, the value function $J^\alpha(x)$ attains its maximal value $J^*(x)=xe^{R_T}$ and the first case is proved.

\vspace{1mm}
\noindent{\bf Step 3}: Now, suppose that
$\wh{\theta}_t$ has non-negative components and ${\xi}_{\lambda}$ defined in \eqref{eq:Mol.6-2} is strictly positive. Then, $\Vert\theta_t\Vert_t>0$ for all $0<t\le T$.  One will try with strategy $(y^*_t=\theta_t \Vert \theta\Vert_\zs{T}^{-1} \rho$ , $v^*_t=0)$ with possible maximal values of $\rho>0$ such that \eqref{eq:Constr.6} is verified. Substituting this particular candidate into this constraint, we observe  that the constraint \eqref{eq:Constr.6} is justified if the following holds
\begin{equation}
\inf_\zs{0\le t\le T} \left\{-\frac{1}{2}u_t^2\rho^2+q_{\beta}u_t\rho+ u_t^2  \Vert \theta\Vert_\zs{T}\rho-\rho K_t\right\}\rho\ge \ln(1-\kappa),
\label{eq:Constr.9}
\end{equation}
where $u_t=\Vert \theta\Vert_t \Vert \theta\Vert_\zs{T}^{-1}\in[0,1]$ and $K_t=(\theta,\xi_{\lambda})_t \Vert \theta\Vert_\zs{T}^{-1}\ge0$. By replacing $K_t$ by $K_\zs{T}$ one gets a stronger requirement
\begin{equation}
\inf_\zs{0\le t\le T} \left\{-\frac{1}{2}u_t^2\rho^2+q_{\beta}u_t\rho+ u_t^2  \Vert \theta\Vert_\zs{T}\rho-K_\zs{T}\rho\right\}\ge \ln(1-\kappa).
\label{eq:Constr.10}
\end{equation}
Let 
$$
g(u,\rho)=-\frac{1}{2}u^2\rho^2+q_{\beta}u\rho+ u^2  \Vert \theta\Vert_\zs{T}\rho-\rho K_T .
$$
Then, $g(0,\rho)=0$ and $g(1, \rho)=-\frac{1}{2}\rho^2+q_{\beta}\rho+ \Vert \theta\Vert_\zs{T}\rho-\rho K_T$. 
Note that $g$ is a strictly decreasing function in $u$ in $[0,1]$. To see this, note that
$$
\partial_u g=\rho(-u\rho+q_\beta+2 \Vert \theta\Vert_\zs{T})<0 \quad \mbox{for all}\quad u\in[0,1]
$$
if $ q_\beta+2\Vert \theta\Vert_\zs{T}<0$, but this is implied by \eqref{eq:Constr.2}. 
So, if one chooses $\rho$ such that 
\begin{equation}
g(1, \rho)=-\frac{1}{2}\rho^2+q_{\beta}\rho+ \Vert \theta\Vert_\zs{T}\rho-\rho K_T =\ln(1-\kappa)
\label{eq:Constr.11}
\end{equation}
then \eqref{eq:Constr.10} is fulfilled. 
Now, equation \eqref{eq:Constr.11} admits a unique positive solution ${\rho}^*_\zs{VaR}$ given by \eqref{eq.Constr.1}. Finally, taking into account the condition $\Vert y\Vert_T\le \sqrt{T}\Vert \sigma\Vert_\zs{T}$ one should choose $\rho=\bar{\rho}^*_\zs{VaR}$ and the proof is completed.\endproof

\vspace{2mm}
Let us now consider Problem \ref{Prob.3} with ES constraint. First, it is useful to rewrite the constraint in the following way
\begin{equation}
\inf_\zs {0\le t\le T}e^{-V_t+(y, \wh{\theta})_t} \ES_\beta({\cal E}_t(y) P_t^\pi(\xi)) \ge (1-\kappa).
\label{eq:ES.1}
\end{equation}
By Lemma \ref{Le.Mol.2}, one sees that \eqref{eq:ES.1} is deduced from  Assumption $(\J)$ and the modified constraint which is independent of jumps 
\begin{equation}
\inf_\zs {0\le t\le T} e^{-V_t+(y, \wh{\theta})_t} \ES_\beta({\cal E}_t(y)) \ge (1-\kappa).
\label{eq:ES.2}
\end{equation}
By Lemma \ref{Le.ES.1} one gets
\begin{equation}
\inf_\zs {0\le t\le T}L_t^{v,y}\ge \ln(1-\kappa) \quad \mbox{and}\quad L_t^{v,y}:=-V_t+(y, \wh{\theta})_t+F_\beta(\Vert y\Vert_t+\vert q_\beta\vert),
\label{eq:ES.3}
\end{equation}
where
\begin{equation}
F_\beta(u)=\ln\left(\frac{1}{\beta}\ov{\Phi}(u)\right)\quad \mbox{and}\quad \ov{\Phi}(u)=1-\Phi(u).
\label{eq:ES.4}
\end{equation}
To formulate the optimal results, we denote by ${\rho}^*_\zs{E}$ the solution of the following equation
\begin{equation}
\Vert \theta\Vert_\zs{T} {\rho}+F_\beta(\rho+\vert q_\beta\vert)-\rho K_T=\ln(1-\kappa).
\label{eq:ES.8}
\end{equation}

\begin{theorem} \label{Th.ES.1}
 Consider Problem \ref{Prob.3} with $\gamma_1=\gamma_2=1$ under assumption $(\J)$. The result in Theorem \ref{Th.VaR.1} still holds if $\rho^*_\zs{VaR}$ is replaced by $\rho^*_\zs{ES}$.
\end{theorem}

\proof Also trying with the strategy $(v^*=0, y^*=\theta_t \Vert \theta\Vert_\zs{T}^{-1} \rho)$ we need to choose the possible maximal value of $\rho$ such that the requirement \eqref{eq:ES.3} is checked. By substituting this particular candidate into $L$ one gets
$$
\inf_\zs {0\le t\le T}L_t^*\ge \psi(u_t,\rho),
$$
where $u_t=\Vert\theta\Vert_t \Vert \theta\Vert_\zs{T}^{-1}\in[0,1]$ and 
\begin{equation}
\psi(u,\rho):=u^2\Vert \theta\Vert_\zs{T}\rho+F_\beta(u\rho+\vert q_\beta\vert)-\rho K_T.
\label{eq:ES.6}
\end{equation}
One will choose $\rho$ such that $\inf_{u\in[0,1],\rho\ge 0} \psi(u,\rho) \ge \ln(1-\kappa)$.
Clearly, $\psi(0,\rho)=0$ and $\psi(1,\rho)=\wh{\psi}(\rho)$, where
\begin{equation}
\wh{\psi}(\rho)=\Vert \theta\Vert_\zs{T} \rho+F_\beta(\rho+\vert q_\beta\vert)-\rho K_T.
\label{eq:ES.7}
\end{equation}
Our aim is to determine a sufficient condition for $\rho$ such that $\wh{\psi}(\rho)$ is the minimum of $\psi(u,\rho)$ on $[0,1]$ and this minimum is equal to $\ln(1-\kappa)$. In other words, one choose $\rho=\rho^*_E$, the solution of equation \eqref{eq:ES.8}.
In order to guarantee that this is in fact the minimum of $\psi(u,\rho)$ on $[0,1]$ one needs to check the sign of the first derivative in $u$. One has, 
$$\partial_u\psi(u,\rho)=2u\Vert \theta\Vert_\zs{T} \rho-\rho\varphi(u\rho+\vert q_\beta\vert)\ov{\Phi}^{-1}(u\rho+\vert q_\beta\vert).
$$
Hence, $\wh{\psi}(\rho)$ is the minimum of $\psi(u,\rho)$ on $[0,1]$ if $\partial_u\psi(u,\rho)<0$ for all $u\in[0,1]$. Using the well-known estimate for the Gaussian integral 
\begin{equation}
(z^{-1}-z^{-3})\varphi(z)< \ov{\Phi}(z)< z^{-1}\varphi(z),\quad z>0
\label{eq:ES.9}
\end{equation}
one gets
$$
\partial_u\psi(u,\rho)<2u\Vert \theta\Vert_\zs{T} \rho-\rho(u\rho+\vert q_\beta\vert)\le \rho(2\Vert \theta\Vert_\zs{T} c-\vert q_\beta\vert)\le 0,
$$
for all $u\in[0,1]$ if $\vert q_\beta\vert\ge 2\Vert \theta\Vert_\zs{T}$. 

Let us verify that equation $\eqref{eq:ES.8}$ has a unique positive solution. To this end, remark that $\wh{\psi}(0)=0$ and $\wh{\psi}(\rho)$ is strictly decreasing if $\vert q_\beta\vert\ge \Vert \theta\Vert_\zs{T}-K_T$. On the other hand, \eqref{eq:ES.9} yields that
$$
\lim_\zs{\rho\to\infty}\wh{\psi}(\rho)=-\infty.
$$
In summery one should take $\rho=\rho^*_E$ defined by equation \eqref{eq:ES.8}. Taking into account the requirement $\Vert y^*\Vert_T\le \sqrt{T}\Vert \sigma\Vert_T$ one gets the same optimal strategy as in Theorem \ref{Th.VaR.1} where $\rho^*_{VaR}$ is replaced by $\rho^*_E$.\endproof

\subsection{The case $0<\gamma_1=\gamma_2=\gamma<1$} 
As in \cite{KluppelbergPergamenchtchikov2009},  we show below that under some mild condition on the model parameters the unconstrained solution in Theorem \ref{Th.Noconstr.2} is still optimal.

Consider first the VaR constraint with non negative jumps. By \eqref{eq:HJB3} one has 
$$
y_t^{*i}=q\wh{\theta}_t^i +q M^i_t,
\quad i=\overline{1,d},
$$
where 
\begin{equation}
M_t^i:=\sum_\zs{j=1}^d  \epsilon_{ij}(t)\int_0^\infty [(1+\pi^{*j}_t z)^{\gamma-1}-1]z\nu^j(\d z).
\label{eq:M}
\end{equation}
Taking into account the above integrals are non positive (since $\gamma<1$), one gets
$y_t^{*i}\le q\wh{\theta}_t^i$ for all $i=\overline{1,d}$. This implies that
$\Vert y^*\Vert_t^2\le q^2\Vert \wh{\theta}\Vert_t^2$. By Lemma \ref{Le:basic},
$$
-V_T^*=\frac{g^q(T)}{\Vert g\Vert ^q_\zs{q,T}+g^q(T)}:=\ln\chi.
$$
Like in the pure diffusion case, the optimal solutions of the constrained and unconstrained problems coincide.
\begin{theorem}\label{Th.Constr.4}
Assume that jumps in assets are non negative and $1-\chi e^{l*(\gamma)}\le \kappa<1,$  where
\begin{equation}
l^*(\gamma):=-q^2 \Vert \wh{\theta}\Vert_T^2+q_{\beta} q \Vert \wh{\theta}\Vert_T.
\label{eq:l}
\end{equation}
Then, under the assumption of Theorem \ref{Th.Noconstr.2} the optimal solution $(y^*,v^*)$ without constraint is also an optimal solution to the corresponding problem with VaR constraint.
\end{theorem}
\proof We need to check the constraint \eqref{eq:Constr.6} for the optimal solution $(y^*_t,v^*)$ of the problem without constraint. For this aim, it suffices to verify that
$$
-\frac{1}{2}\Vert y\Vert_T^2+q_{\beta}\Vert y\Vert_T+\ln \chi\ge \ln(1-\kappa),
$$
but this is an direct consequence of relation $\Vert y^*\Vert_t^2\le q^2\Vert \wh{\theta}\Vert_t^2$ and \eqref{eq:l}. \endproof

Let us consider the problem with ES constraint. To formulate the result, we introduce
\begin{equation}
\wh{M}_t^\theta:=(\wh{\theta}, M)_t\quad \mbox{and}\quad M = (M^1_t,\cdots,M^d_t),
\label{eq:}
\end{equation}
where $M^i$ is defined by \eqref{eq:M}. Observe that $\wh{M}^\theta_t\le 0, \, \forall t\in[0,1]$.
\begin{theorem}\label{Th.Constr.5}
Assume that jumps in assets are non negative, $\vert q_\beta\vert\ge 2\Vert \wh{\theta}\Vert_T$ and the following condition holds 
\begin{equation}
1-\chi \exp\{q\Vert \theta\Vert_\zs{T}^2+F_\beta(q\Vert \wh{\theta}\Vert_T+\vert q_\beta\vert)+\wh{M}_T^\theta\}\le \kappa<1.
\label{eq:M}
\end{equation}
Then, under the assumption of Theorem \ref{Th.Noconstr.2} the optimal solution $(y^*,v^*)$ without constraint is also optimal for the corresponding problem with ES constraint.
\end{theorem}
\proof
We need to check the risk constraint \eqref{eq:ES.3} for strategy $(y^*,v^*)$ defined by system \eqref{eq:HJB3} and \eqref{eq:HJB5}, i.e.$$
-V_t^*+(y^*, \wh{\theta})_t+F_\beta(\Vert y^*\Vert_t+\vert q_\beta\vert) \ge \ln(1-\kappa), \forall t\in[0,1].
$$
Taking into account that $(y, \wh{\theta})_t=\Vert \wh{\theta}\Vert_t+\wh{M}^\theta_t$, $\Vert y^*\Vert_t \le q\Vert \wh{\theta}\Vert_t$ and $F$ is decreasing, we only need to verify that 
\begin{equation}
H(\Vert \wh{\theta}\Vert_t^2)\ge \ln(1-\kappa)\quad \mbox{for all}\quad t\in [0,1],
\label{eq:H1}
\end{equation}
where
$$
H(u):=qu^2+F_\beta(qu+\vert q_\beta\vert)+\wh{M}^\theta_T+\ln\chi.
$$
Using \eqref{eq:ES.9} one observes that 
$$
H'(u)=2qu-q\frac{\varphi(qu+\vert q_\beta\vert)}{\bar{\Phi}(q\Vert \wh{\theta}\Vert_t+\vert q_\beta\vert)}\le 2qu-q(qu+\vert q_\beta\vert)\le 0 \quad \forall u\ge 0
$$
since $\vert q_\beta\vert\ge 2\Vert \wh{\theta}\Vert_T$ by assumption. 
Therefore, $H$ is decreasing in $[0,\infty)$ and \eqref{eq:H1} is verified if $H(\Vert \wh{\theta}\Vert_T)\ge \ln(1-\kappa)$, which can be easily deduced from \eqref{eq:M}.\endproof


\section{The case $0<\gamma_1\neq \gamma_2<1$}\label{General}
We consider the general case where the consumption and bequest functions are different, i.e. $0<\gamma_1\neq \gamma_2<1$. This case is challenging even in pure diffusion models \cite{KluppelbergPergamenchtchikov2009} because the optimal solution without risk constraint does not satisfy the risk constraint. As a matter of fact, the constraint now has a strong impact on the optimal problem.  In particular,  it was shown in \cite {KluppelbergPergamenchtchikov2009} that it is optimal to consume all (at the rate $v^*_t$ which is explicitly determined).
Intuitively, the same result is expected for our present model with jumps. In this section we show that this result is still valid in the presence of jumps by adapting the method used in \cite {KluppelbergPergamenchtchikov2009}. 

We first find an upper bound for the cost function and then try to point out an appropriate control at which the cost function attains that upper bound. First, recall that the cost function is given by 
\begin{equation}
J^\alpha(x):=x^\gamma_1 \int_0^T (v_t e^{-V_t})^{\gamma_1} e^{\gamma_1 R_t} f_1(t,y) \d t+x_2^{\gamma_2}e^{-\gamma_2V_T}e^{\gamma_2 R_T} f_2(T,y),
\label{eq:Diff.1}
\end{equation}
where
\begin{equation*}
f_i(t,y):=\exp\left\{\gamma_i (y,\wh{\theta})_t-\frac{\gamma_i(1-\gamma_i)}{2}\Vert y \Vert _t^2+\sum_{j=1}^d\int_0^t\int_{\bbr}((1+\pi_t^j z)^{\gamma_i}-\gamma_i\pi_t^j z-{\bf 1})\nu^j(\d z)\d t\right\}.
\end{equation*}

\subsection{VaR constraint}
One gets from the risk constraint \eqref{eq:Constr.6} that
$$
-\frac{1}{2}\Vert y\Vert_T^2+q_{\beta}\Vert y\Vert_T-V_T+\Vert\wh{\theta}\Vert_T \Vert y\Vert_T \ge \ln(1-\kappa).
$$
Put $\eta=1-e^{-V_T}$. The above inequality is verified if
\begin{equation}
\Vert y\Vert_T\le \sqrt{(\vert q_{\beta}\vert-\Vert\wh{\theta}\Vert_T)^2 -2\ln(1-\kappa)+2\ln(1-\eta)}-\vert q_{\beta}\vert+\Vert\wh{\theta}\Vert_T:=\rho(\eta),
\label{eq:Diff.2.1}
\end{equation}
for $0\le \eta\le \kappa$. Now, by Holder's inequality and the equality $\int_0^T (v_t e^{-V_t})\d t=1-e^{-V_T}$, one gets
\begin{equation}
J^\alpha(x)\le [x^\gamma_1 \eta^{\gamma_1} \Vert\wh{g}\Vert_\zs{q_1,T}^{q_1}+x_2^{\gamma_2}(1-\eta)^{\gamma_2}\wh{g}_2(T)] \sup_\zs{\Vert y\Vert_T\le \rho(\eta)}\sup_\zs{0\le t\le T} \max(f_1(t,y),f_2(t,y)),
\label{eq:Diff.2}
\end{equation}
where, as in \eqref{eq:HJB.5-1}
$$
\wh{g}_i(t)=e^{\gamma_i R_t}\quad \mbox{and}\quad \Vert\wh{g}\Vert_\zs{q,T}^{q}:=\int_0^{T}\vert\wh{g}(t)\vert^q \d t.
$$

Let us study $\wh{H}_i(x,\eta):=\sup_\zs{\Vert y\Vert_T\le \rho(\eta)}\sup_\zs{0\le t\le T} f_i(t,y)$. Observe first that for $i=1,2$, the function $f_i(t,y)$ attains maximum on the whole admissible set $\cal D$ at $y^{*i}$ satisfying $\Vert y^{*i}\Vert_T \le q_i \Vert \wh{\theta}\Vert_T$.  
By Lemma \ref{Le:compare}, both $f_i(t,y),\, i=1,2$ are concave functions and $f_i(t,y^*)\le \bar{f}_i(q_i\Vert \wh{\theta}\Vert_T),$ where
\begin{equation*}
\bar{f}_i(b):=\exp\left\{\gamma_i \wh{\theta}_T b-\frac{\gamma_i(1-\gamma_i)}{2}b^2\right\},\quad i=1,2.
\end{equation*}
Therefore,
$$
\wh{H}_i(x,\eta)\le \bar{f}_i(\bar{y}_i(\eta)),
$$
where $\bar{y}_i(\eta)=\min(q_i\Vert \wh{\theta}\Vert_T,\rho(\eta))$. It follows that 
$$
J^\alpha(x)\le \max_\zs{i=1,2}\sup_\zs{
0\le \eta\le \kappa} \wh{M}(x,\eta) \bar{f}_i(\bar{y}_i(\eta)),
$$
where
$$
\wh{M}(x,\eta):=x^\gamma_1 \eta^{\gamma_1} \Vert\wh{g}\Vert_\zs{q_1,T}^{q_1}+x_2^{\gamma_2}(1-\eta)^{\gamma_2}\wh{g}_2(T).
$$
\begin{lemma}\label{Le:Diff.1}
Assume that $0<\kappa\le\text{\upshape{argmax}}_\zs{0\le \eta\le 1} \wh{M}(x,\eta)$ and
\begin{equation}
\vert q_{\beta}\vert\ge \Vert\wh{\theta}\Vert_T+ (1-\kappa)^{-1}\Vert\wh{\theta}\Vert_T\max(\gamma_1,\gamma_2)\left(\frac{\partial }{\partial\eta}\ln \wh{M}(x,\eta)\right)^{-1}.
\label{eq:Diff.6}
\end{equation}   
Then, $\wh{M}(x,\kappa)$ is an upper bound of the cost function.
\end{lemma}
\proof
Consider first the case $\rho(0)\le q_i\Vert\wh{\theta}\Vert$. Then $\bar{y}_i(\eta)=\rho(\eta)$ since $\rho$ is decreasing on $[0,\kappa]$. Putting $\wh{G}_i(x,\eta):=\bar{f}_i(\rho(\eta)) \wh{H}_i(x,\eta)$, one deduces that 
$$
\sup_\zs{
0\le \eta\le \kappa} \bar{f}_i(\bar{y}_i(\eta))\wh{M}(x,\eta)=\sup_\zs{
0\le \eta\le \kappa} \bar{f}_i(\rho(\eta)) \wh{M}(x,\eta):=\sup_\zs{
0\le \eta\le \kappa} \wh{G}_i(x,\eta).
$$
Let us study the monotonicity of $\wh{G}_i(x,\eta)$. For this aim, we compute its first derivative
\begin{equation}
\frac{\partial }{\partial \eta}\wh{G}_i(x,\eta)= \gamma_i[\rho'(\eta)\Vert\wh{\theta}\Vert_T-(1-\gamma_i)\rho(\eta)] \bar{f}_i(\rho(\eta)) \wh{M}(x,\eta)+\bar{f}_i(\rho(\eta)) \frac{\partial }{\partial \eta}\wh{M}(x,\eta).
\label{eq:Diff.3}
\end{equation}
Note that $\wh{M}(x,\eta)$ is a concave function, which has first positive derivative and negative second derivative on $[0,\kappa]$ provided that $\kappa\in[0, \text{argmax}_\zs{0\le \eta\le 1} \wh{M}(x,\eta)]$. 
Therefore, $\frac{\partial }{\partial_\eta}\wh{G}_i(x,\eta)\ge 0$ if 
\begin{equation}
 \frac{\partial }{\partial\eta}\ln \wh{M}(x,\eta)\ge  \gamma_i \Vert\wh{\theta}\Vert_T\sup_\zs{\eta\in[0,\kappa]}\vert\rho'(\eta)\vert.
\label{eq:Diff.4}
\end{equation}
On the other hand, one gets from \eqref{eq:Diff.2.1} that
$$
\rho'(\eta)=-(1-\eta)^{-1}[\vert q_{\beta}\vert-\Vert\wh{\theta}\Vert_T)^2 -2\ln(1-\kappa)+2\ln(1-\eta)]^{-1/2}.
$$
Then, 
$$
\sup_\zs{0\le\eta\le \kappa}\vert \rho'(\eta)\vert
 \le (1-\kappa)^{-1}(\vert q_{\beta}\vert-\Vert\wh{\theta}\Vert_T)^{-1}.
$$
So, \eqref{eq:Diff.4} holds if
\begin{equation}
\frac{\partial }{\partial\eta}\ln \wh{M}(x,\eta)\ge \gamma_i \Vert\wh{\theta}\Vert_T (1-\kappa)^{-1}(\vert q_{\beta}\vert-\Vert\wh{\theta}\Vert_T)^{-1}, \quad i=1,2. 
\label{eq:Diff.5}
\end{equation}
Observe that \eqref{eq:Diff.5} is equivalent to \eqref{eq:Diff.6}.                                     
Under \eqref{eq:Diff.6}, $\wh{G}_i(x,\cdot)$ is increasing, which implies that $\sup_\zs{
0\le \eta\le \kappa} \wh{G}_i(x,\eta)=\wh{G}_i(x,\kappa)=\wh{M}_i(x,\kappa)$ since $\bar{f}_i(x,\kappa)=1.$

\vspace{2mm}
Suppose now that $q_i\Vert\wh{\theta}\Vert<\rho(0)$. Recall that $\rho$ is decreasing with $\rho(\kappa)=0$. There exists $\eta_i\in[0,\kappa]$ such that $ q_i\Vert\wh{\theta}\Vert=\rho(\kappa_i)$. Now,
$$
\sup_\zs{
0\le \eta\le \kappa_i} \bar{f}_i(\bar{y}_i(\eta))\wh{M}(x,\eta)=\sup_\zs{
0\le \eta\le \kappa_i}\bar{f}_i(\rho(\eta_i))\wh{M}(x,\eta)=\bar{f}_i(\rho(\eta_i))\wh{M}(x,\eta)=\wh{G}_i(x,\eta_i).
$$
On the other hand, observe that $\bar{y}_i(\eta)=\rho(\eta)$ if $\eta\in[\kappa_i,\kappa]$. As already shown above,
$$
\sup_\zs{
\kappa_i\le \eta\le \kappa} \bar{f}_i(\bar{y}_i(\eta))\wh{M}(x,\eta)=\wh{G}_i(x,\kappa).
$$
As $\wh{G}_i(x,\cdot)$ is increasing one concludes that $\sup_\zs{
0\le \eta\le \kappa} \wh{G}_i(x,\eta) =\wh{G}_i(x,\kappa)=\wh{M}(x,\kappa)$.
Hence, $\wh{M}(x,\kappa)$ is always an upper bound of the cost function.\endproof

\begin{theorem}\label{Th:Diff.1}
Under the assumptions of Lemma \ref{Le:Diff.1}, $(y^*=0,v^*)$ is the optimal solution for the problem with VaR risk constraint, where
\begin{equation}
v^*_t=\dot{V}^*_t=\frac{\kappa \wh{g}_1^{q_1}(t)}{\Vert \wh{g}_1\Vert_\zs{q_1,T}^{q_1}-\kappa\Vert\wh{g}_1\Vert^{q_1}_\zs{q_1,t}(t)}.
\label{eq:Diff.7}
\end{equation}
\end{theorem}
\proof
We need to find an control at which the cost function attains the upper bound $\wh{M}(x,\kappa)$. Clearly, we should choose $v$ such that
$$
\int_0^T (v_t e^{-V_t})^{\gamma_1} \wh{g}_1(t)=(1-e^{-V_T})^{\gamma_1} \Vert \wh{g}_1\Vert_\zs{q_1,T}\quad\mbox{and}\quad V_T=-\ln(1-\kappa).
$$
For this aim, we solves the differential equation on $[0,T]$
$$
\dot{V}_t e^{-V_t}=\frac{\kappa}{\Vert \wh{g}_1\Vert_\zs{q_1,T}^{q_1}} \wh{g}^{q_1}_1(t),\quad V_0
=0.
$$
The last differential equation admits solution 
$$
V^*_t=-\ln\left(1-\frac{\kappa}{\Vert \wh{g}_1\Vert_\zs{q_1,T}^{q_1}} \Vert\wh{g}_1\Vert^{q_1}_\zs{q_1,t}(t)\right),
$$
which gives the optimal consumption rate defined in \eqref{eq:Diff.7}. \endproof

\subsection{ES constraint}
Consider Problem \ref{Prob.3} with ES constraint. An upper bound for the cost function is given by the following.
\begin{lemma}\label{Le:Diff.3}
Assume that $0<\kappa\le\text{\upshape{argmax}}_\zs{0\le \eta\le 1} \wh{M}(x,\eta)$ and
\begin{equation}
\vert q_{\beta}\vert\ge 2\Vert\wh{\theta}\Vert_T+ (1-\kappa)\Vert\wh{\theta}\Vert_T\min(\gamma_1,\gamma_2)\left(\frac{\partial }{\partial\eta}\ln \wh{M}(x,\eta)\right)^{-1}.
\label{eq:Diff.8}
\end{equation}   
Then, $\wh{M}(x,\kappa)$ is an upper bound of the cost function.
\end{lemma}
\proof Recall that $V_T=\ln(1-\eta)$. First the risk constraint \eqref{eq:ES.3} implies that 
\begin{equation}
\Vert y\Vert_t\Vert \wh{\theta}\Vert_T+F_\beta(\Vert y\Vert_t+\vert q_\beta\vert)\ge \ln(1-\kappa)-\ln(1-\eta),\quad \forall t\in[0,1].
\label{eq:Diff.9}
\end{equation}
It is easy to check that the function 
$
\wh{\psi}(\rho):=\Vert \wh{\theta}\Vert_T \rho+F_\beta(\rho+\vert q_\beta\vert)
$
is strictly decreasing if $\vert q_\beta\vert\ge \Vert \wh{\theta}\Vert_T$. So, \eqref{eq:Diff.9} is checked if we have
$$
\wh{\psi}(\Vert y\Vert_T)\ge \ln(1-\kappa)-\ln(1-\eta).
$$ Note that $\wh{\psi}(\Vert y\Vert_T)\le \wh{\psi}(0)=F_\beta(q_\beta\vert)=0 $. Therefore, there exists a solution $\wh{\rho}:=\wh{\rho}(\eta)$ for equation 
\begin{equation}
\wh{\psi}(\rho)=\ln(1-\kappa)-\ln(1-\eta),\quad 0\le \eta\le \kappa.
\label{eq:Diff.10}
\end{equation}
and $\wh{\eta}$ is strictly decreasing since $\wh{\psi}(\rho) $ is decreasing, which implies $\wh{\eta}\ge \wh{0}$. Now, taking derivative of two sides \eqref{eq:Diff.10} one gets
$$
\wh{\rho}'(\eta)\left[\Vert y\Vert_T-\frac{\varphi(\wh{\eta}+\vert q_\beta\vert)}{\bar{\Phi}(\wh{\eta}+\vert q_\beta\vert)}\right]=\frac{1}{1-\eta}.
$$
Observe that the expression in the square brackets is bounded by $\Vert y\Vert_T-\vert q_\beta\vert$. It follows that
$$
\vert\wh{\rho}'(\eta)\le\frac{1}{(1-\eta)(\vert q_\beta\vert-\Vert y\Vert_T)},\quad \eta\in[0,\kappa].
$$ 
At this stage, the analysis as Lemma \ref{Le:Diff.1} can be applied to show that $\wh{M}(x,\kappa)$ is an upper bound of the cost function. \endproof

We have proved that under the ES constraint, it is optimal to consume all. 
\begin{theorem}\label{Th:Diff.2}
Under the assumptions of Lemma \ref{Le:Diff.1}, $(y^*=0,v^*)$ where $v^*$ is given by \eqref{eq:Diff.7}, is the optimal solution for the problem with ES risk constraint.
\end{theorem}

Theorems \ref{Th:Diff.1} and \ref{Th:Diff.2} show that the presence of a dynamical risk constraint has an undesired effect that the investor whose portfolio constitutes in both consumption and investment should optimally consume all. Thus, our results suggest considering the utility maximization problem (optimal investment) and the optimal consumption problem separately. Remark that \cite{Aitsahalia09} considers the unconstrained consumption problem in a similar jump-diffusion setting. In particular, by exploiting differences in the Brownian risk of the asset returns that lies in the orthogonal space, the authors show that optimal policy can be obtained by focusing on controlling the exposure to the jump risk.
\subsection{When consumption is not possible}
Let us consider the following utility maximization problem 
$
\sup_{\pi}\E [U(X_T^\pi)],
$
where the wealth process is given by
\begin{equation}
\d X_t^\pi =X_t^\pi(r_t+y'_t\theta_t)\d t+X_t^\pi y_t^{'}\d W_t+X_\zs{t^{-}}^\pi \int_{\bbr^d}\pi_t^{'} z \wt{J}(\d z\times \d t), \quad X_0^\pi=x>0.
\label{eq:nocons.1}
\end{equation}
One then deduces the HJB equation
\begin{equation}
\partial_t u(t,x)+\sup_\zs{\pi}  {\cal A} ^\pi u(t,x)=0, \quad u(T,x)=x^\gamma,
\label{eq:HJB0}
\end{equation}
where the generator ${\cal A}^\alpha$ is defined by as in \eqref{eq:Oper} with $v=0$:
\begin{align}
{\cal A}^\pi u(t,x)&=x(r_t+y'_t \wh{\theta}_t)\partial_xu(t,x)+\frac{1}{2}x^2 y_t y'_t\partial_{xx}^2u(t,x)\notag\\
&+\sum_\zs{j=1}^d \int_\bbr (u(t,x+x\pi^j_tz)-u(t,x)-x\pi^j_tz\partial_xu(t,x))\nu^j(\d z),
\label{eq:}
\end{align}
We try to find a solution of the form $u(t,x)=\rho x^\gamma$, and $
u_x=\gamma\rho x^{\gamma-1},\, u_\zs{xx}=\gamma (\gamma-1)\rho x^{\gamma-2},
$
where $\rho$ is a $t$-function to be determined.
Substituting these formulas into \eqref{eq:HJB} we obtain 
\begin{align}
\rho'(t)+\rho(t)\sup_\zs{\pi} \left\{\gamma (r_t+y'_t \wh{\theta}_t)+\frac{1}{2} y_t^2 \gamma (\gamma-1)
+\sum_\zs{j=1}^d \K^j(\pi^j_t) \right\}=0,
\label{eq:HJBnc}
\end{align}
where $\K^j$ defined in \eqref{eq:K}. Thus, we get the same necessary optimal condition as in \eqref{eq:HJB3}, i.e.
\begin{equation}
\wh{\theta}_t^i +(\gamma-1)y_t^i+\sum_\zs{j=1}^d  \epsilon_{ij}(t) Q^{j}(\pi^j_t)=0,\quad y^i\ge 0, \,i=\overline{1,d},
\label{eq:HJB30}
\end{equation}
where $Q^{j}$ defined in \eqref{eq:HJB4}. Direct argument leads to the optimal solution for the unconstrained problem.
\begin{proposition}\label{Pro.Noconsumpt.1}
Assume that $y^*\ge 0$ is a solution of system \eqref{eq:HJB30} and let $h^*$ be the corresponding supremum in \eqref{eq:HJBnc}. Then, the optimal rule for the unconstrained problem is given by $y^*$ and the value function is $e^{\int_t^T h^*(s)\d s} x^{\gamma}$. The wealth process is given by
\begin{equation}
\d X_t^* =X_t^*(r_t+y_t^*\theta_t)\d t+X_t^* y_t^*\d W_t+X_\zs{t^{-}}^* \int_{\bbr^d}\pi_t^* z \d\wt{J}(\d z\times \d t), \quad X_0^*=x>0.
\label{eq:noconsumpt}
\end{equation}
\end{proposition}
\proof The conclusion follows a similar argument as in Theorem \ref{Th.Noconstr.2} with the remark that $e^{\int_t^T h^*(s)\d s}$ is the solution to the ordinary differential equation $\rho'(t)+\rho(t) h^*(t)=0$.\endproof

Let us turn to the constrained problem $\sup_{\pi}\E [U(X_T^\pi)]$
under VaR/ES constraint. The following is just a direct consequence of Theorems \ref{Th.Constr.4} and \ref{Th.Constr.5}.
\begin{proposition}\label{Pro.Noconsumpt.2}
Assume that jumps in the assets are non negative and condition \eqref{eq:l} holds. Then the unconstrained solution $y^*$ in Proposition \ref{Pro.Noconsumpt.1} is still optimal solution to the utility maximization problem with VaR constraint. The same conclusion is still true for ES constraint when \eqref{eq:l} is replaced with condition \eqref{eq:M}.
\end{proposition}
Proposition \eqref{Pro.Noconsumpt.2} provides sufficient conditions so that the risk constraints VaR/ES are not active. When these conditions do not hold, it could be possible to incorporate the constraint into the HJB equation. This makes the problem more attractive but more challenging to solve, which we do not pursue it here. In fact, the challenge lies in the fact that the risk constraint is impossible to transformed into an explicit constraint on strategies as in the pure diffusion case due to the presence of jumps. Nevertheless, the HJB equation with investment constraints can be solve numerically as in \cite{EmmerKlupperbergKorn01} with an approximation on jumps sizes. In \cite{DuffiePan01}, the authors provide an analytical method (applying an analytical Fourier-transform) for computing value at risk, and other risk measures that allows for fat-tailed and skewed return distributions.
\section{Negative jumps}\label{Negative}
We examine in this section the effect of negative jumps. In fact, the regulator should be more conservative if  negative jumps probably happen. We show below that in that case, a slightly stricter constrained depending on the probability of having negative jumps in the risky assets during the investment horizon $[0,T]$ can be imposed to make the previous analysis still valid.

First, note that we still have $M_t^i\le 0 $ for all $i=\overline{1,d}$ even in the presence of negative jumps. Let us begin by  examining the VaR constraint inequality \eqref{eq:Constr.4} by introducing
\begin{equation}
\wh{P}_t^\pi(\xi):=\frac{1}{\Vert y\Vert_t}\sum_{j=1}^d \sum_{k=1}^{N_t^j}  \ln(1+\pi^j_\zs{\tau_k^{j{-}}} \xi^j_k).
\label{eq:Nega1}
\end{equation}
We want to find a lower bound for $q_{\beta}({\cal E}_t(y) P_t^\pi(\xi))$. The latter can be written as
 $\exp\{-\frac{1}{2}\Vert y\Vert_t^2+\wh{q}_{\beta}(t)\Vert y\Vert_t\}$, where
$\wh{q}_{\beta}(t):=q_{\beta}(Z_t+ \wh{P}_t^\pi(\xi))$ and $Z_t$ is a ${\cal F}_t$ measurable standard normal variable independent of $\wh{P}_t^\pi(\xi)$. Denote by $\varepsilon_t:=\P(A_t)$, where
\begin{equation}
A_t:=\{ \text{there are at least one negative jump in the asset prices in}\, [0,t]\}
\label{eq:Nega2}
\end{equation}
Now, the assumption that the jump parts of the risky assets are independent leads to the following elementary property.

\begin{lemma} For any $0\le t\le T$, we have
$$
\epsilon_t=\prod_{i=1}^d (1-e^{-\lambda_i t})\int_{-1}^0 F^i (\d z).
$$
\end{lemma}

Obviously, $\varepsilon_t \nearrow \varepsilon_T$ as $t\to T$. We try to estimate $\wh{q}_{\beta}(t)$ respect to $\varepsilon_T$. From the quantile definition, we have
\begin{equation}
\beta=\P(Z_t+ \wh{P}_t^\pi(\xi)\le \wh{q}_{\beta}(t))
\le 
\P(Z_t+\wh{P}_t^\pi(\xi)\le \wh{q}_{\beta}(t), A_t^c)+\P (A_t).
\label{eq:Nega3}
\end{equation}
On $A_t^c:=\Omega\backslash A_t$ ( which is independent of $Z_t$), $\wh{P}_t^\pi(\xi)$ is non negative. It follows that
$$
\beta\le \P(Z_t\le \wh{q}_{\beta}(t)) \P(A_t^c)+\P(A_t)=\P(Z_t\le \wh{q}_{\beta}(t))(1-\varepsilon_t)+\varepsilon_t,
$$
or $\wh{q}_{\beta}(t)\ge q_\zs{\wh{\beta}(\varepsilon_t)}$, where
\begin{equation}
\wh{\beta}(\varepsilon):=\frac{\beta-\varepsilon}{1-\varepsilon}.
\label{eq:Nega4}
\end{equation}
Clearly, $\wh{\beta}'(\varepsilon)=[2\varepsilon-(1+\beta)]/(1-\varepsilon)^2<0$ if $\varepsilon\le (\beta+1)/2$. Thus, $\wh{\beta}(\varepsilon)$ is decreasing down to $\wh{\beta}_T:=\wh{\beta}(\varepsilon_T)$ if $0\le \varepsilon_T\le (\beta+1)/2,$ which implies that $\wh{q}_{\beta}(t)\ge q_\zs{\wh{\beta}_T}$.
We then deduce that the risk constraint \eqref{eq:Constr.4} is checked if 
\begin{equation}
\inf_\zs{0\le t\le T} \left\{-\frac{1}{2}\Vert y\Vert_t^2+q_{\wh{\beta}_T}\Vert y\Vert_t-V_t+(y,\wh{\theta})_t\right\}\ge \ln(1-\kappa).
\label{eq:Neg.5}
\end{equation}
Hence, we need to replace $q_{\beta}$ with $q_{\wh{\beta}_T}$ in the analysis before.

Remark that using the modified confidence level $\wh{\beta}_T\le \beta$ means that the investor' portfolio is more strictly regulatory. In general, during a normal day-life period of time $\varepsilon_T=0$ even though the asset prices are allowed to go down but in a continuous way (predictable). Note that the analysis given in the previous sections can be obtained by sending $\varepsilon_T$ to zero. 
%
%

\vspace{2mm}

Let us now consider Problem \ref{Prob.3} with ES constraint defined by \eqref{eq:ES.1}. Along the lines in the proof of Lemma \ref{Le.Mol.2} and by the positivity of ${\cal E}_t(y) P_t^\pi(\xi)$ one gets
$$
\ES_\beta({\cal E}_t(y) P_t^\pi(\xi))=\frac{1}{\beta}\int_0^{\beta} q_\delta({\cal E}_t(y) P_t^\pi(\xi)) \d \delta\ge \frac{1}{\beta}\int_{\varepsilon_T}^{\beta} q_\delta({\cal E}_t(y) P_t^\pi(\xi)) \d \delta.
$$
As above, 
$$
q_\delta({\cal E}_t(y) P_t^\pi(\xi))\ge q_\zs{\wh{\delta}_t}({\cal E}_t(y))\ge q_\zs{\wh{\delta}_T}({\cal E}_t(y)),
$$
where 
$$
\wh{\delta}_t:=\wh{\delta}(\epsilon_t)=\frac{\delta-\varepsilon_t}{1-\varepsilon_t}.
$$
After changing variable we obtain
$$
\ES_\beta({\cal E}_t(y) P_t^\pi(\xi))\ge \frac{1}{\beta}\int_{\varepsilon_T}^{\beta} q_\zs{\wh{\delta}_T}({\cal E}_t(y)) \d \delta=\frac{\beta-\varepsilon_T}{\beta} \frac{1}{\wh{\beta}}\int_{0}^{\wh{\beta}} q_\zs{\wh{\delta}_T}({\cal E}_t(y)) \d \wh{\delta}_T.
$$
In other words,
$$
\ES_\beta({\cal E}_t(y) P_t^\pi(\xi))\ge \frac{\beta-\varepsilon_T}{\beta} \ES_\zs{\wh{\beta}_T}({\cal E}_t(y)),
$$
where $\wh{\beta}_T$ is defined in \eqref{eq:Nega4}. Now, by Lemma \ref{Le.ES.1}
$$
\ES_\zs{\wh{\beta}_T}({\cal E}_t(y))=\frac{1}{\wh{\beta}_T}(1-\Phi(\vert q_{\wh{\beta}_T}\vert +\Vert y\Vert_t))=F_{\wh{\beta}_T}(\Vert y\Vert_t+\vert q_{\wh{\beta}_T}\vert).
$$
Thus, in the presence of negative jumps, we need to replace the function $F_\beta$ in the risk constraint \eqref{eq:ES.3} with $\wh{F}_{\wh{\beta}_T}$, defined by 
\begin{equation}
\wh{F}_{\wh{\beta}_T}(u):=\frac{\beta-\varepsilon_T}{\beta} F_{\wh{\beta}_T}(u).
\label{eq:Nega.7}
\end{equation}
Then, the optimal policy can be obtained after a similar procedure. 

\vspace{2mm}

In summary, we have proved the following main results.

\begin{theorem}\label{Th:Nega.1}
Assume that negative jumps during the considered horizon $[0,T]$ take place with probability $\varepsilon_T<\beta$. Then, all results obtained in the previous sections for VaR constraint are still valid if we replace the level $\beta$ with $\wh{\beta}_T$ defined by \eqref{eq:Nega4} in the corresponding risk constraint. For the problem with ES constraint, we have the same result by replacing $F_\beta$ with $\wh{F}_{\wh{\beta}_T}$ given in \eqref{eq:Nega.7}.
\end{theorem}

\section{Concluding remark}
We studied the problem of optimal investment and consumption under VaR and ES risk constraints focusing on deterministic strategies. When jumps in asset are non negative, the approach in \cite{KluppelbergPergamenchtchikov2009} can be applied to get the optimal solution among a subset of admissible strategies obtained by ignoring jumps in the constraint but with the same confidence level. In particular, we showed that under some mild condition on the model parameters, the unconstrained solution is still optimal if two identical power utility functions are used. For different utility functions, the impact of constraint is dramatic and it is optimal for the investor to consume all. When negative jumps probably happen, the regulator should be more conservative to impose a  slightly stricter constrained depending on the probability of having negative jumps in the risky assets during the investment horizon, to ensure that the analysis for the case of non negative jump is still valid.

It should be noticed that random strategies can be considered, but, to make the HJB approach still valid, it is necessary to modify the definition of quantile or use a relative VaR/ES constraint \cite{Pirvu07}. We also plan to extend the present paper to general Levy's models. In such cases, approximating small jumps needs to be studied and the problem of stability may be interesting to investigate as in \cite{EmmerKlupperbergKorn01,liu2003dynamic}. 

\vspace{1.5cm}

\noindent{\bf \Large Appendix: Auxiliary results}

\setcounter{section}{0}
\renewcommand{\thesection}{\Alph{section}}
\section{Quantile and expected shorfall}\label{App}
\begin{definition}[Lower quantile]\label{Def.0-1}
For any random variable $Y$ and $\beta\in(0,1)$, the lower $\beta$-quantile of $Y$ is the number defined by
\begin{equation}
q_\beta(Y)=\inf\{ u: \P(Y\le u)\ge \beta\}.
\label{eq:Mol.12}
\end{equation}
\end{definition}
The following is useful to provide an explicit form for the optimal solution.

\begin{lemma}\label{Le.Mol.1}
Let $q_\beta$ be the lower $\beta$-quantile of the standard normal distribution and ${\cal E}_t(y)$ be the stochastic exponential defined in \eqref{eq:Mol.6-1}. Then,
\begin{equation}
q_\beta({\cal E}_t(y))=\exp\left\{-\frac{1}{2}\Vert y\Vert_t^2+q_\zs{\beta}\Vert y\Vert_t\right\}.
\label{eq:Mol.6-20}
\end{equation}

\end{lemma}
\proof It follows directly from the definition of ${\cal E}_t(y)$ and the linearity of lower quantile.\endproof

\begin{definition}[Expected Shorfall]\label{Def.2}
 For any random variable $Y$ and $\beta\in(0,1)$, the expected shorfall at $\beta$-quantile of $Y$ is the real number $\ES_\beta(Y)$ defined by 
\begin{equation}
\ES_\beta(Y)=\E(Y\vert Y\le q_\beta(Y))
\label{eq:Mol.15}
\end{equation}
for some random variable $Y$.
\end{definition}
Again, the following simple result is useful to get the optimal solution in an explicit form.
\begin{lemma}\label{Le.ES.1}
Let $\phi$ be the standard normal distribution function and ${\cal E}_t(y)$ be the stochastic exponential defined in \eqref{eq:Mol.6-1}. For any $\beta\in(0,1)$, we have
\begin{equation}
\ES_\beta( {\cal E}_t(y))=\frac{1}{\beta}(1-\Phi(\vert q_\beta\vert +\Vert y\Vert_t)),
\label{eq:Mol.15-1}
\end{equation}
where $q_\beta$ is the lower $\beta$-quantile of a standard normal random variable.
\end{lemma}
\proof The definition \eqref{eq:Mol.15} implies that
$$
\ES_\beta( {\cal E}_t(y))=e^{-\frac{1}{2}\Vert y\Vert_t^2}\E(e^{\Vert y\Vert_t Z}\vert Z\le q_\beta)=\frac{e^{-\frac{1}{2}\Vert y\Vert_t^2}}{\P(Z\le q_\beta)} \E(e^{\Vert y\Vert_t Z}{\bf 1}_\zs{\{Z\le q_\beta\}}),
$$
where $Z\sim N(0,1)$. Direct calculus provides that 
$\P(Z\le q_\beta)= \beta$ and 
$$
e^{-\frac{1}{2}\Vert y\Vert_t^2}\E(e^{\Vert y\Vert_t Z}{\bf 1}_\zs{\{Z\le q_\beta\}})=\Phi(\vert q_\beta\vert +\Vert y\Vert_t),
$$
and the conclusion follows.\endproof

Risk of two portfolios can be compared using the following lemma.
\begin{lemma}\label{Le.Mol.2} Let $Y,Z$ be two random variables satisfying $Y\le Z$ a.s. Then, for any $\beta \in (0,1)$,
\begin{equation}
q_\beta(Y)\le q_\beta(Z)\quad \mbox{and}\quad \ES_\beta(Y)\le \ES_\beta(Z).
\label{eq:Mol.16}
\end{equation}
\end{lemma}
\proof
For any $u\in \bbr$,  one has $\P(Y\le u) \ge \P(Z\le u)$ since $Y\le Z$ almost surely. Then, 
$$
\{ u: \P(Z\le u)\ge \beta\}\subset \{ u: \P(Y\le u)\ge \beta\}
$$
for any $\beta\in(0,1)$ and hence $q_\beta(Y)\le q_\beta(Y)$. Let us prove the last inequality in  \eqref{eq:Mol.16}. Clearly, 
$$
\P(Y\le q_\beta(Y))=\beta\quad \mbox{and}\quad q_\beta(Y)=F_Y(\beta),
$$
where $F_Y$ is the distribution function of $Y$. The same representation can be also obtained for $Z$. Now, by definition,
$$
\ES_\beta(Y)=\E(Y\vert Y\le q_\beta(Y))=\frac{1}{\P(Y\le q_\beta(Y))}\E(Y{\bf 1}_\zs{\{ Y\le q_\beta(Y)\}})=\frac{1}{\beta}\int_{-\infty}^{q_\beta(Y)} y F_Y(\d y).
$$
By changing variable $\delta=F_Y(y)\rightarrow y=F_Y^{-1}(\delta)=q_\delta(Y)$ one obtains 
$$
\ES_\beta(Y)=\frac{1}{\beta}\int_0^{\beta} q_\delta(Y) \d \delta \quad\mbox{and}\quad
\ES_\beta(Z)=\frac{1}{\beta}\int_0^{\beta} q_\delta(Z) \d \delta
$$
and the conclusion follows by the first inequality.\endproof
\section{Geometric L\'evy martingale}
\begin{lemma}\label{Le:Levy}
Let $a:[0,T]\times \bbr \to \bbr$ be a function satisfying 
$$
\E\Big[\exp\Big\{\int_0^T\int_{\bbr}(e^{a(t,z)}-1) \nu (\d z) \d t\Big\}\Big]<\infty.
$$
Then, the process $X_t$ defined by $ \d X_t=X_t^{-} \int_{\bbr}(e^{a(t,z)}-1) \wt{J}(\d z\times \d t)$ is a martingale and 
$$
 \E\Big[\exp\Big\{\int_0^T\int_{\bbr}{a(t,z)}{J}(\d z\times \d t) \Big\}\Big]=\E\Big[\exp\Big\{\int_0^T\int_{\bbr}(e^{a(t,z)}-1) \nu (\d z) \d t\Big\}\Big]<\infty.
$$
\end{lemma}
\proof See exercise 1.6 in \cite{OksendalSulem07}. \endproof
\section{Exponential of optimal consumption rate}
\begin{lemma}\label{Le:basic} For $v_t^*$ defined in \eqref{eq:HJB5} we have
\begin{equation}
e^{-V_T^*}=\frac{g^q(T)}{\Vert g\Vert ^q_\zs{q,T}+g^q(T)}
\label{eq:}
\end{equation}
\end{lemma}
\proof It seems that a direct verification using \eqref{eq:HJB5} is technically hard. We may proceed as follows. Provided $\pi=\pi^*$ is an optimal portfolio, we need to choose $v$ such that the cost function is maximal, i.e.
$$
\max_\zs{v}\int_0^T v^{\gamma}_te^{-\gamma V_t} g(t) \d t+e^{-\gamma V_T} g(T),
$$
where $g$ is defined by \eqref{eq:rho}. This variation problem can be solved in two steps. First, By Holder's inequality the above formula is bounded by
$$
\int_0^T v^{\gamma}_te^{-\gamma V_t} g(t) \d t \le \int_0^T v_te^{- V_t} \d t \Vert g\Vert ^q_\zs{q,T}+e^{-\gamma V_T} g(T).
$$
The equality happens when $v^{\gamma}_te^{-\gamma V_t}$ and $g(t)$ are linearly independent in $L^1$, i.e., $v_te^{-V_t}=b g^q(t)$ a.s. on $[0,1]$, where $b$ is a positive constant. It follows that
$$
1-e^{-V_T}=\int_0^T v_te^{-V_t} \d t=b \Vert g\Vert ^q_\zs{q,T},
$$
which implies that $e^{-V_T}=1-b \Vert g\Vert ^q_\zs{q,T}$ and the cost function is now given by
$$
f(b)=b^\gamma \Vert g\Vert ^q_\zs{q,T}+(1-b \Vert g\Vert ^q_\zs{q,T})^\gamma g(T).
$$
It remains to maximize $f(b)$ by choosing an appropriate $b>0$. As $f$ is concave, its maximum attains at the zero point of the first derivative $f'(b^*)=0$, where
$
b^*=[{g^q(T)+\Vert g\Vert ^q_\zs{q,T}}]^{-1}.
$
Thus, $$e^{-V_T^*}=1-b^* \Vert g\Vert ^q_\zs{q,T}=\frac{g^q(T)}{g^q(T)+\Vert g\Vert ^q_\zs{q,T}}.\endproof
$$

\section{Proof of Theorem \ref{Th.Noconstr.1}}\label{App:1}
By \eqref{eq:Mol.7}, one has 
\begin{equation}
\E X_t^\alpha= xe^{R_t-V_t+(y,\wh{\theta})_t} \E{\cal E}_t(y) \E P_t^\pi(\xi)=xe^{R_t-V_t+(y,\wh{\theta})_t}\E P_t^\pi(\xi)
\label{eq:Noconstr.01}
\end{equation}
since $\E{\cal E}_t(y)=1$. Taking into account the independency of the terms in \eqref{eq:Mol.9} one obtains
\begin{equation}
\E P_t^\pi(\xi)=\exp\left\{\sum_{j=1}^d\int_0^t\int_{\bbr}\pi^j_s z^j \nu^j(\d z^j) \d t\right\}=\exp\left\{\int_0^t\pi_s  {\xi}_{\lambda} \d s\right\}.
\label{eq:Noconstr.0}
\end{equation}
Therefore, 
\begin{equation}
\E X_t^\alpha= xe^{R_t-V_t+(y,\wh{\theta})_t+(\pi, {\xi}_{\lambda})_t}=xe^{R_t-V_t+(y,{\theta})_t}=xe^{R_t-V_t+(\pi,\mu-r{\bf 1})_t},
\label{eq:Noconstr.1}
\end{equation}
which is bounded by $xe^{R_T} e^{-V_t+\Vert\pi\Vert_\zs{T}\Vert \mu-r{\bf 1}\Vert_\zs{T}}.$
We then deduce that 
\begin{align*}
 J^\alpha(x)&=\int_0^T \E X^\alpha_t v_t \d t+\E X^\alpha_T \\
&\le xe^{R_T+\Vert\pi\Vert_\zs{T}\Vert \mu-r{\bf 1}\Vert_\zs{T}}\left(\int_0^T e^{-V_t}v_t \d t+e^{-V_T}\right)
=xe^{R_T+\Vert\pi\Vert_\zs{T}\Vert \mu-r{\bf 1}\Vert_\zs{T}}.
\end{align*}
Hence, $xe^{R_T+T\Vert \mu-r{\bf 1}\Vert_\zs{T}}$ is an upper bound of $J^\alpha(x)$ since $\Vert\pi\Vert_\zs{T}\le \sqrt{T}$. 
Now, if $\Vert \mu-r{\bf 1}\Vert_\zs{T}=0$, the upper bound is attainable for any admissible strategy $\pi^*$ with $\pi^{*j}_t\in[0,1]$ and $v^*_t=0$. In the contrary case if $\Vert \mu-r{\bf 1}\Vert_\zs{T}>0$, then the upper bound is attained for $\pi^*_t=(\mu_t-r_t{\bf 1})\Vert \mu-r{\bf 1}\Vert_\zs{T}^{-1}\sqrt{T}$ and $v^*_t=0$.  \endproof

\section{Proof of Lemma \ref{Le:compare}}\label{App:com}
Recall that $\pi_t^*$ and $\bar{\pi}_t^*$ are respectively maximal points of 
$$
G(t,\pi)=\gamma r_t+\gamma(\mu_t-r_t)\pi +\frac{1}{2}\sigma_t^2 \pi^2 \gamma (\gamma-1)+\K(\pi_t)
$$
and 
$$
\bar{G}(t,\pi):=\gamma r_t+\gamma(\mu_t-r_t)\pi+\frac{1}{2}\sigma_t^2 \pi^2 \gamma (\gamma-1).
$$
In other words, they are solutions to equations $\kappa:=\partial_\pi G(t,\pi)=0$ and $\bar{\kappa}(\pi)=0$, where
$$
\kappa(\pi):=\gamma[\mu_t-r_t +(\gamma-1)\sigma^2_t \pi+ \int_\bbr [(1+\pi z)^{\gamma-1}-1]z\nu(\d z)] 
$$
and
$$
\bar{\kappa}(\pi):=\mu_t-r_t+(\gamma-1)\sigma^2_t \pi.
$$
Clearly, $\kappa (\pi)\le \bar{\kappa}(\pi),\,\forall \pi\in [0,1]$ and those two functions are decreasing. One obtains that $\pi^*\le \bar{\pi}^*$. Moreover, as both $G$ and $\bar{G}$ are concave in $[0,1]$ one has $$G^*:=\max G(t,\pi)\le \bar{G}^*_t:=\max \bar{G}(t,\pi),$$
which in turn leads to the comparison $\rho(t)\le \bar{\rho}(t)$ (see the defintion in  \eqref{eq:HJB5}). Taking into the negative sign of $\gamma-1$ one gets $v^*_t=\rho(t)^{1/(\gamma-1)}\ge \bar{v}^*_t=\bar{\rho}(t)^{1/(\gamma-1)}$. \endproof
\bibliographystyle{plain}
\bibliography{Jump_Opt_Risk}

\end{document}